\documentclass{WileyMSP-template}
\usepackage{epsfig,amssymb,amsmath,mathrsfs,amsfonts,color,amsthm,graphicx,float,color,bm,url}
\usepackage[colorlinks=true]{hyperref}
\usepackage{makecell}
\usepackage{booktabs}
\usepackage{multirow}
\usepackage{graphicx,color}
\usepackage[table]{xcolor}
\usepackage{afterpage}

\usepackage{array}
\newcolumntype{P}[1]{>{\centering\arraybackslash}p{#1}}
\newcolumntype{M}[1]{>{\centering\arraybackslash}m{#1}}



\usepackage[active]{srcltx}

\usepackage{subfigure}

\usepackage{qcircuit}
 \def\map#1{\mathcal #1}

\def\<{\langle}\def\>{\rangle}
\def\bb{\langle\!\langle}\def\kk{\rangle\!\rangle}
\def\Tr{\operatorname{Tr}}
\def\:{\hbox{\bf
    :}}

\def\R{\mathbb R}

\def\L{\mathcal{L}}

\def\Supp{\mathsf{Supp}}  
\def\spc#1{\mathcal{#1}}

\DeclareMathOperator*{\argmax}{arg\,max}
\DeclareMathOperator*{\argmin}{arg\,min}

\def\Conv{\mathsf{Conv}}

\def\seq{\mathsf{Seq}}
\def\strat{\mathsf{Strat}}
\def\rank{\mathsf{rank}}
\def\dim{\mathsf{dim}}

\newcommand{\h}[1]{\mathcal{H}_{#1}}
\newcommand{\dyad}[1]{|#1\>\<#1|}

\newtheorem{theo}{{Theorem}}
\newtheorem{defi}{{Definition}}
\newtheorem{lemma}{{Lemma}}

\newtheorem{cor}{{Corollary}}

\usepackage[ruled,vlined]{algorithm2e}

\begin{document}

\rhead{\includegraphics[width=2.5cm]{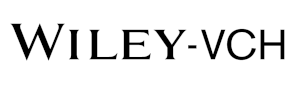}}

\title{Fully-Optimized Quantum Metrology: Framework, Tools, and Applications}

\maketitle

\author{Qiushi Liu}
\author{Zihao Hu}
\author{Haidong Yuan*}
\author{Yuxiang Yang*}

\begin{affiliations}
Qiushi Liu\\
QICI Quantum Information and Computation Initiative, Department of Computer Science, The University of Hong Kong, Pokfulam Road, Hong Kong, China\\
 
Zihao Hu\\ 
Department of Mechanical and Automation Engineering, The Chinese University of Hong Kong, Shatin, Hong Kong, China\\

Prof.~Haidong Yuan\\ 
Department of Mechanical and Automation Engineering, The Chinese University of Hong Kong, Shatin, Hong Kong, China\\
hdyuan@mae.cuhk.edu.hk\\

Prof.~Yuxiang Yang\\
QICI Quantum Information and Computation Initiative, Department of Computer Science, The University of Hong Kong, Pokfulam Road, Hong Kong, China\\
yuxiang@cs.hku.hk

\end{affiliations}


\begin{abstract}
This tutorial introduces a systematic approach for addressing the key question of quantum metrology: For a generic task of sensing an unknown parameter, what is the ultimate precision given a constrained set of admissible strategies. The approach outputs the maximal attainable precision (in terms of the maximum of quantum Fisher information) as a semidefinite program and optimal strategies as feasible solutions thereof. Remarkably, the approach can identify the optimal precision for different sets of strategies, including parallel, sequential, quantum SWITCH-enhanced, causally superposed, and generic indefinite-causal-order strategies. The tutorial consists of a pedagogic introduction to the background and mathematical tools of optimal quantum metrology, a detailed derivation of the main approach, and various concrete examples. As shown in the tutorial, applications of the approach include, but are not limited to, strict hierarchy of strategies in noisy quantum metrology, memory effect in non-Markovian metrology, and designing optimal strategies. Compared with traditional approaches, the approach here yields the exact value of the optimal precision, offering more accurate criteria for experiments and practical applications. It also allows for the comparison between conventional strategies and the recently discovered causally-indefinite strategies, serving as a powerful tool for exploring this new area of quantum metrology.

\end{abstract}

\tableofcontents

\section{Introduction}

Quantum metrology \cite{giovannetti2004quantum,Giovannetti2006PRL} is the science of boosting the accuracy of sensing by quantum entanglement and coherence, featuring a broad spectrum of applications including enhanced gravitational wave detection \cite{Schnabel2010,LIGO19PRL}, quantum sensor networks \cite{Komar2014}, and quantum radars \cite{Barzanjeh15PRL}. As quantum technologies advent, quantum metrology will benefit from the finer control of quantum probes and higher versatility of metrology protocols. Therefore, given a sensing task, it is natural to ask what is the ultimate precision of quantum metrology and how this precision can possibly be achieved.

\begin{figure}[!hbtp]
    \centering
    \includegraphics[width=0.7\linewidth]{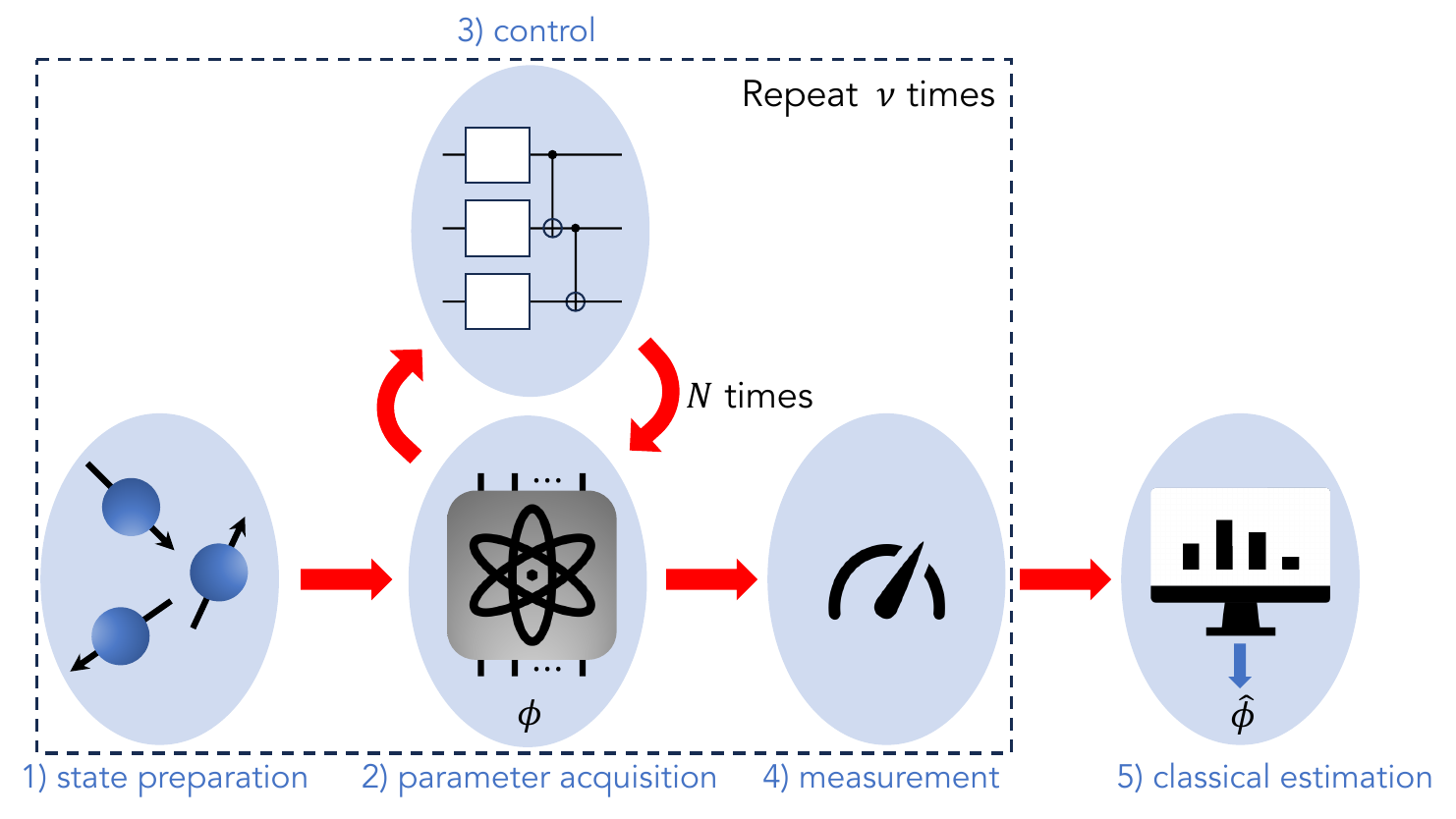}
    \caption{{\bf A general setup of quantum metrology.} The general procedure of estimating $\phi$ from an unknown physical process consists of 5 steps: $1)$ preparing a probe state, $2)$ letting the probe interact with the unknown process to acquire information on $\phi$, $3)$ applying control operations on the probe, $4)$ measuring the probe, and $5)$ processing the experimental data to generate an estimate $\hat{\phi}$. }
    \label{fig:flowchart}
\end{figure}

As shown in Figure~\ref{fig:flowchart}, preparing signal-sensitive probe states, applying suitable intermediate control, and performing the right measurement all lead to a higher precision potentially. Optimizing them in a unified fashion to determine the ultimate precision, on the other hand, is a daunting task traditionally.  
As shown in Figure~\ref{fig:flowchart}, the probe may interact with the signal source coherently for $N$ times. In the limit of $N\gg 1$, depending on the task, it is asymptotically optimal to prepare spin-squeezed probe states \cite{Kitagawa1993PRA}, incorporate control operations \cite{Yuan2015PRL,Yuan2016PRL} or apply error correction \cite{Duer2014PRL,Arrad14PRL,Kessler2014PRL,ozeri2013heisenberg,Sekatski2017quantummetrology,Demkowicz-Dobrza2017PRX,Zhou2018,Zhou2021PRXQ}, achieving a mean squared error (MSE) that scales as $1/N$ (the standard quantum limit) or $1/N^2$ (the Heisenberg limit). In contrast, in the practically relevant regime of finite $N$, little is known on the dependence of the optimal precision and the optimal strategy on $N$, and the optimality of existing strategies holds only in the asymptotic sense.

Another related key question of quantum metrology is the hierarchy of strategies.
Traditionally, it is a fundamental question \cite{Demkowicz-Dobrzanski14PRL} to compare parallel strategies, where the $N$ queries to the signal are applied simultaneously on a large entangled probe state, and sequential strategies, where the $N$ queries are applied one after another. 
It was shown that sequential strategies are strictly superior for estimation of multiple parameters of a unitary \cite{Yuan2016PRL}, while the two families of strategies have asymptotically equal performance in the limit of large $N$ \cite{Zhou2021PRXQ,kurdzialek2023using}. 
However, for finite $N$, the technical tool for comparing different strategies was missing. 

Going beyond these strategies of fixed causality,
it was also found in recent works that causally indefinite quantum information processing involving a quantum SWITCH, where the orders of signal queries are placed in a quantum superposition, can further enhance the precision of metrology even beyond the quantum limit \cite{Zhao2020PRL,Yin2023NP}. There has been a rapidly growing interest in quantum metrology with indefinite causal order (ICO) \cite{Frey2019,mukhopadhyay2018superposition,Zhao2020PRL,Chapeau-Blondeau2021PRA,Goldberg23PRR,Yin2023NP,An24PRAnoisy}. On the other hand, most literature considered toy models without optimizing the strategies, for an optimization tool was also missing in the ICO setting.

Given a generic quantum sensing task, this tutorial covers a systematic approach of determining the strictly maximal precision as well as strategies that attain this precision. Via the approach, the maximal precision (in terms of the quantum Fisher information) is expressed as a semidefinite program (SDP) and thus can be numerically evaluated. The approach also identifies the optimal performances under different strategies, including not only sequential and parallel strategies but also causally indefinite strategies such as those that involve a quantum SWITCH.

This tutorial not only offers a comprehensive introduction to the optimal quantum metrology approach in Refs.~\cite{Yang2019PRL,Altherr2021PRL,Liu23PRLoptimal}, but also provides various new examples and detailed case studies. For example, via our approach we find a strict hierarchy between causally indefinite, sequential and parallel strategies exists for a finite $N$ in both cases of asymptotic channel estimation following the Heisenberg limit and the standard quantum limit. We also show a strict advantage of indefinite causal order over any predetermined causal order in estimating two different noisy channels. In non-Markovian quantum metrology, we compare the performances between parallel ``feedforward'' strategies and sequential feedback ones, with or without control. Moreover, we discuss both SDP and variational circuits for designing optimal strategies.

 The remaining part of the tutorial is organized as follows: In Section \ref{sec:background}, we start with basic notions and useful tools. We introduce the problem formulation in quantum metrology, the quantum comb formalism for characterizing quantum processes, and the conversion of quantum combs into circuits. In Section \ref{sec:strategy sets}, we formally define several quantum strategy sets under different causal constraints in the comb formalism. The main results are presented in Section \ref{sec:optimal metrology}, including an approach to both the optimal estimation precision (Theorem \ref{thm_qfi_sdp}) and an optimal strategy (Algorithm \ref{alg_optimal_strategy}) that achieves it, as well as detailed proofs. We conclude Section \ref{sec:optimal metrology} with a comparison between our framework and conventional approaches. In Section \ref{sec:applications} we apply our theoretical approach to plenty of models in the estimation of both Markovian and non-Markovian quantum processes, and also compare the metrological performance together with circuit complexity between SDP and two alternative variational approaches for designing optimal strategies. Conclusions and outlook are presented in Section \ref{sec:conclusions}.

\section{Background} \label{sec:background}
In this section, we prepare basic notions of quantum metrology and the mathematical tools for deriving the main result. Readers familiar with quantum metrology and (or) quantum comb formalism may skip the corresponding part(s).
In this tutorial, we denote by $\L(\h{})$ linear operators on a Hilbert space $\h{}$, and by $\L(\h{1},\h{2})$ linear transformations from $\h{1}$ to $\h{2}$.
When there is no ambiguity we will write $A\ge 0$ for a positive semi-definite operator $A$, and $\Tr_i A$ denotes the partial trace of $A$ over the subspace $\h{i}$. 

\subsection{Introduction to quantum metrology}
\threesubsection{Quantum parameter estimation}
Parameter estimation is a fundamental task in statistics. Given a biased coin, one may toss it for many times, and count the frequencies of the head and the tail to infer its bias. In the quantum regime, the coin is replaced by quantum states. As a simple example, suppose we want to estimate the energy gap $\Delta\epsilon:=\epsilon_1-\epsilon_0$ of a two-level system with Hamiltonian $H=\epsilon_0|0\>\<0|+\epsilon_1|1\>\<1|$. We can prepare the state $|+\>:=(1/\sqrt{2})(|0\>+|1\>)$ and let it evolve under $H$ for fixed time $\tau$. The final state is then $|+_{\Delta\epsilon}\>=(1/\sqrt{2})(|0\>+e^{-i\Delta\epsilon\tau}|1\>)$ up to an irrelevant global phase. Measuring it in the basis $\{|\pm\>:=(1/\sqrt{2})(|0\>\pm|1\>)\}$ yields the outcome ``+'' with probability $P(+)=\cos^2(\Delta\epsilon\tau/2)$. Therefore, we may repeat the experiment for $\nu\gg1$ times to estimate $P(+)$ and, subsequently, $\Delta\epsilon$.

The task of quantum parameter estimation is to estimate an unknown parameter $\phi$,
by properly measuring copies of a state $\rho_\phi$ that depends on the unknown parameter.
Naturally, two most important questions are $i)$ what is the ultimate precision and $ii)$ how to achieve it. When $\phi\in\R$ is a single parameter, these questions are addressed by the quantum Cram\'{e}r-Rao bound \cite{helstrom1976quantum,holevo2011probabilistic} on the mean-squared error of any unbiased estimation strategy\footnote{When the estimation is biased, the error does not decrease with $\nu$. Therefore, it is common to focus on unbiased estimation.}
\begin{align}\label{qcrb}
\delta^2_\phi\ge\frac{1}{\nu J(\rho_\phi)}.
\end{align}
Here $J(\rho_\phi)$ is the \emph{quantum Fisher information} (QFI) of the state family $\{\rho_\phi\}$ at $\phi$, and $\nu$ is the number of times that the experiment is repeated. Crucially, the above bound is achievable in the large-$\nu$ asymptotic regime or, put in another way, when one has a large number of copies of $\rho_\phi$. 
Therefore, the ultimate precision (i.e., the maximum of $1/\delta^2_\phi$) is proportional to the QFI $J(\rho_\phi)$ of the state.
Note that, unless specified, we focus on the case of single-parameter estimation. When there is more than one parameters unknown, the attainability of Eq.~(\ref{qcrb}) is not guaranteed, but the QFI still serves as a key measure of precision. We will comment on this toward the end of this tutorial.

Thanks to the quantum Cram\'{e}r-Rao bound, we can now focus on the QFI and regard it as the score function for the precision.
There are multiple methods of evaluating the QFI (see, for instance, the review paper \cite{liu2020Quantum}). Here it is enough to focus on the following formula:
\begin{lemma}[QFI of quantum states \cite{Fujiwara2008}]
The QFI of a parametrized family of quantum states $\{\rho_\phi\}$ can be evaluated as
\begin{align}\label{def_state_qfi}
J(\rho_\phi)=4\min_{|\Psi_\phi\>:\Tr_A|\Psi_\phi\>\<\Psi_\phi|=\rho_\phi}\<\dot{\Psi}_\phi|\dot{\Psi}_\phi\>.
\end{align}
Here the minimization is over any parametrized pure state family $\{|\Psi_\phi\>\}$ that purifies $\{\rho_\phi\}$, and $|\dot{\Psi}_\phi\>$ denotes the partial derivative of $|\Psi_\phi\>$ with respect to $\phi$.
\end{lemma}
Note that two purifications of $\rho_\phi$ are related by a unitary on the environment, which can be $\phi$-dependent.

An immediate consequence of the above lemma is that the QFI is monotonically non-increasing under discarding subsystems. Consider a generic parametrized state family $\{\rho_\phi\}$ with $\rho_\phi\in\L(\h{1}\otimes\h{2})$, and $\{\sigma_\phi\}$ with $\sigma_\phi=\Tr_2\rho_\phi$ where $\Tr_2$ denotes the partial trace over $\h{2}$. Then, in Eq.~(\ref{def_state_qfi}), the domain of the minimization for $\sigma_\phi$ contains the domain of the minimization for $\rho_\phi$, since a purification of $\rho_\phi$ is always a purification of $\sigma_\phi$. As a consequence, the result of the minimization can potentially be smaller. It is also immediate from Eq.~(\ref{def_comb_qfi}) that the QFI remains unchanged if one applies any $\phi$-independent isometry to $\rho_\phi$. Combining these two observations, we conclude that the QFI satisfies the following data processing inequality:
\begin{cor}[Monotonicity of the QFI]\label{cor_monotone_qfi}
Let $\{\rho_\phi\}$ be a generic parametrized family of states on $\h{1}$ and $\map{C}:\L(\h{1})\to\L(\h{2})$ be an arbitrary quantum channel (i.e., a completely-positive trace-preserving (CPTP) linear map) that does not depend on $\phi$. Then, the following inequality holds:
\begin{align}\label{monotonicity_qfi}
J(\map{C}(\rho_\phi))\le J(\rho_\phi).
\end{align}
\end{cor}

To achieve the quantum Cram\'{e}r-Rao bound (\ref{monotonicity_qfi}), one has to measure $\rho_\phi$ with a suitable measurement. By Corollary \ref{cor_monotone_qfi}, the QFI will at best remain unchanged after the measurement (regarding any measurement as an entanglement breaking channel). There is indeed always a measurement that preserves the QFI for single-parameter estimation: 
Let $L_\phi$ be the symmetric logarithmic derivative (SLD) operator, defined as the solution to $\dot{\rho}_\phi=(1/2)(L_\phi\rho_\phi+\rho_\phi L_\phi)$. It is immediate that the Hermitian operator $(1/2)(L_\phi+L_\phi^\dag)$ is also a solution. We can thus assume w.o.l.g. $L_\phi$ to be a (Hermitian) observable with eigenbasis $\{|L_{\phi,j}\>\}$. It can be shown that measuring $\rho_\phi$ in the basis $\{|L_{\phi,j}\>\}$ yields a probability distribution 
\begin{align}
    \left\{q_{\phi,j}:=\<L_{\phi,j}|\rho_\phi|L_{\phi,j}\>\right\}
\end{align}
whose (classical) Fisher information, defined as
\begin{align} \label{eq:CFI}
    J_{\rm cl}\left(\{q_{\phi,j}\}\right):=\sum_j\frac{(\dot{q}_{\phi,j})^2}{q_{\phi,j}},
\end{align}
achieves the QFI
\begin{align}
    J_{\rm cl}\left(\{q_{\phi,j}\}\right)=\max_{\map{M}:{\rm measurement}}J(\map{M}(\rho_\phi))=J(\rho_\phi).
\end{align}
It is important to note that sometimes the SLD $L_\phi$ may depend on the unknown parameter $\phi$ and thus the optimal measurement cannot be applied without prior information on $\phi$. From this perspective, the quantum Cram\'{e}r-Rao bound (\ref{qcrb}) holds only locally in a small neighborhood of the true value of the parameter to estimate. For global estimation of $\phi$ in a non-vanishing range, we may adopt a two-step procedure, where one first runs full tomography on a small portion of $\rho_\phi$'s copies to get a rough estimate on $\phi$ and then measure the remaining copies with the SLD corresponding to the estimate. This approach attains the quantum Cram\'{e}r-Rao bound (\ref{qcrb}) in general; see, for instance, Ref.~\cite[Section 9]{yang2019attaining} for more details.

\threesubsection{Quantum metrology tasks} Quite often we need to estimate a parameter encoded dynamically in a physical process instead of in a state, which is the typical setting of quantum metrology, as illustrated in Figure~\ref{fig:flowchart}.
Quantum parameter estimation, which we just discussed, corresponds to Figure~\ref{fig:flowchart} with 1) state preparation being fixed and 3) control being trivial. For general quantum metrology, it is more complex to find the ultimate precision, as both the state preparation and the control are flexible for optimization. 

The role of control during the parameter acquisition stage can be crucial. 
Let us consider the task of estimating $\phi$ encoded in the Hamiltonian of a two-level system $H_\phi=\cos(\phi) X+\sin(\phi) Z$ with $X,Z$ being the Pauli operators. Intuitively, longer interrogation time suggests better precision in the absence of noise.
However, the maximal QFI obtained by optimizing the initial probe state and not applying any control is $4\sin^2(\tau)$, which may go down with the total interrogation time $\tau$. This can be remedied by a suitable feedback control \cite{Yuan2015PRL}.
Therefore, to achieve optimal metrology, it is crucial to optimize the design of the entire strategy including both the probe state and the control, which is a challenging job in general.

In this tutorial, we consider a generic scenario, where one prepares a probe state and sends it through the physical process. During the interrogation, it is allowed to apply for $N-1$ times on the probe state suitable control operations. It is assumed that either the operations are very fast (compared to the evolution to estimate), as in the case of dynamical decoupling \cite{viola1998Dynamical,viola1999Dynamical}, or one can pause the interrogation while doing the control operations. Either way, in this picture, the physical process is an $N$-step process that incorporates the control operations and the state preparation. 

The course of estimating $\phi$ consists of three major procedures: preparing a probe state, letting it interact with $\map{C}_\phi$, and measuring the final state with a proper measurement. We treat the first two in a unified way and describe them by a \emph{strategy}. Note that we do not worry about the measurement due to the attainability of the quantum Cram\'{e}r-Rao bound (\ref{qcrb}).
The goal of quantum metrology is to find over a set of allowed strategies an optimal one that outputs a state with maximal QFI.
\begin{defi}[Metrology tasks]\label{defi-metro-task}
A (quantum) metrology task is specified by a pair $(\map{C}_\phi,\strat)$, where $\map{C}_\phi$ is an $N$-step physical process carrying the parameter of interest, and $\strat$ denotes the set of strategies that one can apply.
\end{defi}
Note that we abused the notation a bit: for a task one needs to specify the whole set of processes $\{\map{C}_\phi\}$. Next, we introduce a mathematical framework for optimizing quantum metrology.

\subsection{The comb formalism of quantum processes}
This subsection introduces a mathematical tool for faithfully describing our $N$-step physical processes and, consequently, both $\map{C}_\phi$ and the strategies.

\threesubsection{Quantum combs}
The well-known Choi-Jamiołkowski isomorphism \cite{jamiolkowski1972linear,choi1975completely} states that a generic one-step quantum process, i.e., a quantum channel $\map{C}$ from an input Hilbert space $\h{1}$ to an output Hilbert space $\h{2}$, can be faithfully described by an operator on $\h{1}\otimes\h{2}$, named the \emph{Choi operator} $C$, satisfying the constraint $\Tr_2 C=I_1$, $C\ge 0$. Crucially, the converse also holds: any $C\in\L(\h{1}\otimes\h{2})$ satisfying these constraints correspond to a quantum channel $\map{C}:\h{1}\to\h{2}$. 

\begin{figure}[H]
    \centering
    \includegraphics[width=0.7\linewidth]{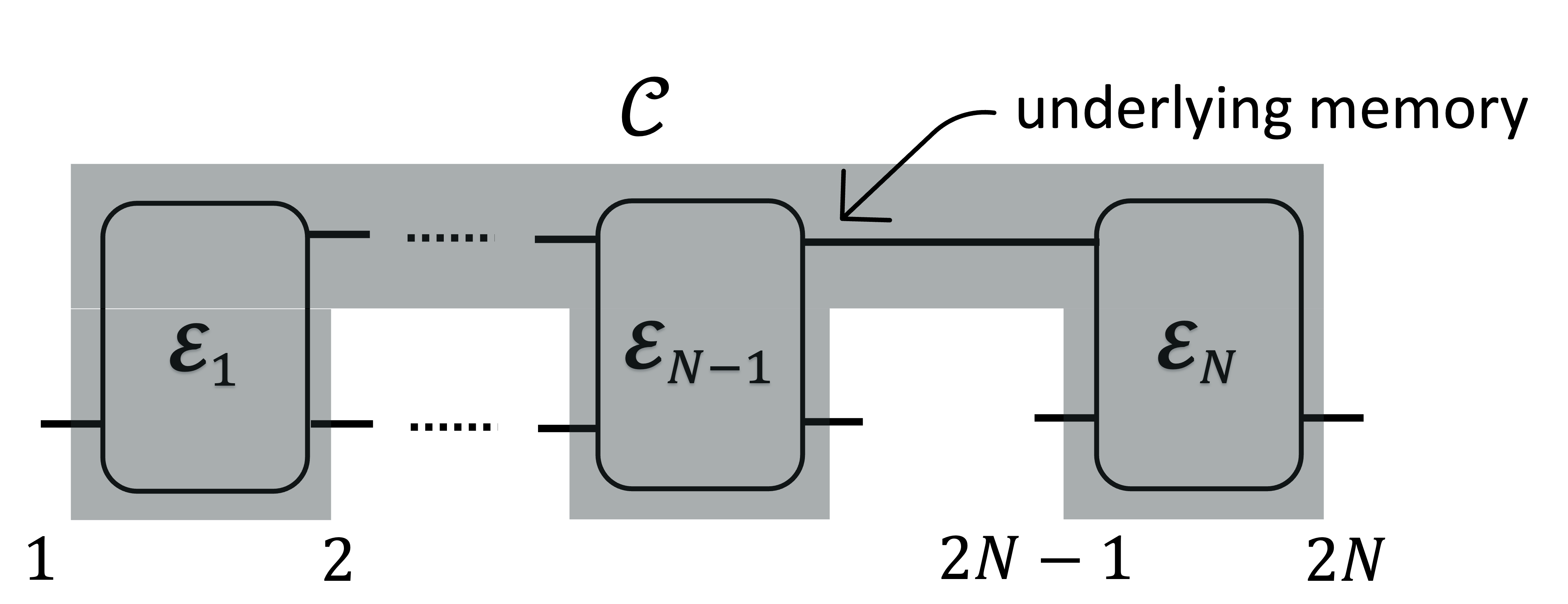}
    \caption{A generic $N$-step quantum process with definite causal order.}
    \label{fig:gen_comb}
\end{figure}

This result extends to $N$-step physical processes \cite{gutoski2007toward,Chiribella2008PRL,Chiribella2009PRA}. As shown in Figure~\ref{fig:gen_comb}, an $N$-step quantum process $\map{C}$ (with definite causal order) is captured by its \emph{quantum comb}, which is a Choi operator $C\in\L(\otimes_{i=1}^{2N}\h{i})$ satisfying the following constraints:
\begin{equation}\label{comb-def}
\begin{aligned}
C&\ge 0,\qquad  
_{2N,\dots,2i,2i-1} C = _{2N,\dots,2i}C,\ i=1,\dots,N, 
\end{aligned}
\end{equation}
where we have used the notation
\begin{align}
    _{i} C:=\frac{I_{\h{i}}}{d_{i}}\otimes\Tr_{\h{i}} C,
\end{align} 
where $d_i:=\dim(\h{i})$ is the dimension of the Hilbert space $\h{i}$. The first constraint ensures the complete positivity (CP) of the process, and the other constraints guarantee that the process always yields legitimate processes, states, or probability distributions when interlaced with other processes. It is noteworthy that the subscripts $2N,2N-1,\dots,1$ appear in a descending order due to the causality constraints of the process (e.g., information flows into $\h{3}$ cannot influence the input to $\h{1}$). We will see processes where some of these constraints are lifted later. On the other hand, given a generic quantum process $\map{C}$, its (quantum) comb can be determined by inserting half of the (unnormalized) maximally entangled state on the joint input subspace:
\begin{align}\label{def_map_to_operator}
C=\map{C}\otimes\map{I}(|I\kk \bb I|),
\end{align}
where $|I\kk:=\sum_n|n\>|n\>\in\tilde{\h{}}_{\rm in}\otimes\h{{\rm in}}$ for an orthonormal basis $\{|n\>\}$ of $\h{{\rm in}}:=\otimes_{i=1}^N\h{2i-1}$ and $\tilde{\h{}}_{\rm in}\simeq\h{{\rm in}}$. Note that we will use the ``double-ket" notation $|A\kk:=(A\otimes I)|I\kk=(I\otimes A^T)|I\kk$ for any matrix $A$. The correspondence between some common quantum processes and their Choi operators is summarized in Table~\ref{tab:choi}.

 \begin{table}[h]
     \centering
     \begin{tabular}{c|c|c}
        Quantum process  & Definition & Comb \\
        \hline
        state preparation  & $\rho$ & $\rho$ \\
        \hline
        (partial) trace of $\h{i}$  & $\Tr_{i}$ & $I_{i}$\\
        \hline
        quantum channel & $\sum_i K_i(\cdot)K_i^\dag$ for Kraus operators $\{K_i\}_i$ & $\sum_i|K_i\kk\bb K_i|$\\
        \hline
        identity channel & $(\cdot)\to(\cdot)$ & $|I\kk\bb I|$\\
        \hline
         measure and prepare & $\sum_i\Tr[(\cdot)M_i]\sigma_i$ for a POVM $\{M_i\}$ and density matrices $\sigma_i$  & $\sum_i   (M_i)^T\otimes\rho_i$\\ 
        \hline
         \makecell{neutralization of\\ the input-output pair $(\h{i-1},\h{i})$} & \makecell{inserting a maximally mixed state into $\h{i-1}$\\ and discarding the output of $\h{i}$}
         & $\frac{I_{i-1,i}}{d_{i-1}}$
         \\
         \hline
         \makecell{neutralization of\\ the output-input pair $(\h{i},\h{i+1})$} & $\frac{I_{i+1}}{d_{i+1}}\Tr_i[(\cdot)]$ & $\frac{I_{i,i+1}}{d_{i+1}}$\\
         \hline
     \end{tabular}
     \caption{Combs of some quantum processes}
     \label{tab:choi}
 \end{table}

\threesubsection{Composition of combs}
Next, we use the quantum comb framework to describe the intervention on a physical process.
Consider the combs $A\in\L(\h{1}\otimes\h{2})$ and $B\in\L(\h{2}\otimes\h{3})$ of two generic physical processes $\map{A}$ and $\map{B}$. $\h{2}$ may be further decomposed into several subspaces $\otimes_{j}\h{2,j}$, and the actions of $\map{A}$ and $\map{B}$ on these subspaces are complementary, i.e., if $\h{2,j}$ is an output of $\map{A}$ then it should be an input for $\map{B}$. 
It is then possible to compose $\map{A}$ with $\map{B}$ to form a new process, whose comb is given by the \emph{link product} of $A$ and $B$:
\begin{align}\label{def_link_product}
A\ast B:=\Tr_{2}[(A^{T_2}\otimes I_3)(I_1\otimes B)],
\end{align}
where $A^{T_2}$ denotes the partial transpose of $A$ on $\h{2}$. 
As a sanity check, consider a simple example of composing two quantum channels $\map{A}:\h{1}\to\h{2}$ and $\map{B}:\h{2}\to\h{3}$, with Choi operators $A\in\L(\h{1}\otimes\h{2})$ and $B\in\L(\h{2}\otimes\h{3})$. Recall that the action of a quantum channel on any input state is given by
\begin{align}\label{def_operator_to_map}
\map{C}(\rho)=\Tr_{R}[(\rho_R^{T}\otimes I)C],
\end{align}
where $\h{R}$ is a copy of the input Hilbert space of $\map{C}$. Using Eq.~(\ref{def_operator_to_map}) twice, the output state of the composed channel $\map{B}\circ\map{A}$ can be expressed in terms of their Choi operators $A$ and $B$ as
\begin{equation}
    \begin{aligned}
    \map{B}\circ\map{A}(\rho) &= \Tr_{2}[(\Tr_{1}[(\rho_1^{T}\otimes I_2)A]^{T}\otimes I_3)B] \\
    &= \Tr_1[(\rho_1^{T}\otimes I_3)\Tr_{2}[(A^{T_2}\otimes I_3)(I_1 \otimes B)]],
    \end{aligned}
\end{equation}
Comparing with Eq.~(\ref{def_operator_to_map}), we see that the Choi operator of $\map{B}\circ\map{A}$ is $\Tr_{2}[(A^{T_2}\otimes I_3)(I_1 \otimes B)]$, which is exactly the link product $A\ast B$ defined in Eq.~(\ref{def_link_product}). Thus the Choi operator of the composed quantum channel is the link product of the Choi operators of its components.

\threesubsection{Necessity and sufficiency of the condition (\ref{comb-def}).}
Now we consider composing channels $\map{E_1}:\h{1}\to\h{2,2'}$ and $\map{E_2}:\h{2',3}\to\h{4}$ with Choi operators $E_1\in\L(\h{1}\otimes\h{2}\otimes\h{2'})$ and $E_2\in\L(\h{2'}\otimes\h{3}\otimes\h{4})$, where $\h{i,j}$ stands for $\h{i}\otimes\h{j}$. If we draw the quantum circuit and rearrange the wires representing the input and output quantum systems, the composed map $\map{E_2}\circ\map{E_1}$ can be reshaped into the form of a comb (i.e., a multi-step quantum channel with an inaccessible memory). We thus refer to the Choi operator $E_1\ast E_2$ as a quantum two-comb. It is straightforward to generalize this definition to a quantum $N$-comb defined as $C_N= E_1\ast E_2 \ast \cdots \ast E_N$, with $E_i$ being the Choi operator of the $i$-th channel, as shown in Figure~\ref{fig:gen_comb}. There is no ambiguity in the definition of $C_N$ since associativity is automatically satisfied by the construction of quantum combs and link product is commutative up to a reordering of Hilbert spaces.

Here we show how the constraints in Eq.~(\ref{comb-def}) can be derived from the standard quantum circuit model (see also Ref.~\cite{Chiribella2009PRA}). We will label the input and output Hilbert spaces of $\map{E_i}$ as $\h{2i-1,(2i-1)'}$ and $\h{2i,(2i)'}$. ($\h{(2i-1)'}$ and $\h{(2i)'}$ are underlying memory subspaces; $\h{1'}$ and $\h{(2N)'}$ are both trivial.) We now show that the trace preserving (TP) conditions for $\map{E_i}, 1\le i\le N$ result in the recursive constraints in Eq.~(\ref{comb-def}). 

Given a quantum comb $C$, we first show that $_{2N}C= _{2N,2N-1}C$. Applying the neutralization of the input-output pair $(\h{2N-1},\h{2N})$ to $C=E_1\ast\cdots\ast E_N$ (cf.~Table~\ref{tab:choi}), we get 
$\frac{I_{2N-1}}{d_{2N-1}}\ast I_{2N}\ast C = \frac{I_{2N-1}}{d_{2N-1}}\ast (I_{2N}\ast C) =\frac{I_{2N-1}}{d_{2N-1}}\ast(E_1\ast\cdots\ast E_{N-1})\ast(I_{2N}\ast E_N)$. By the TP condition for the last channel $\map{E}_N$, we have $I_{2N} \ast E_N=\Tr_{2N}E_N=I_{2N-1,(2N-1)'}=I_{2N-1}\ast I_{(2N-1)'}$ for $\map{E}_N:\h{2N-1,(2N-1)'}\to\h{2N}$. Therefore, we have $\frac{I_{2N-1}}{d_{2N-1}}\ast I_{2N}\ast C = \frac{I_{2N-1}}{d_{2N-1}}\ast(E_1\ast\cdots\ast E_{N-1})\ast(I_{2N-1}\ast I_{(2N-1)'})= (E_1\ast\cdots\ast E_{N-1})\ast(\frac{I_{2N-1}}{d_{2N-1}}\ast I_{2N-1})\ast I_{(2N-1)'}=C_{N-1}$, where $C_{N-1}:=E_1\ast\cdots\ast E_{N-1}\ast I_{(2N-1)'}$.
On the other hand, $I_{2N}\ast C$ results in the Choi operator of the $N$-step process with $\h{2N}$ traced out. We have $I_{2N}\ast C = (E_1\ast\cdots E_{2N-1})\ast(I_{2N}\ast E_{2N})=C_{N-1}\ast I_{2N-1} = C_{N-1}\otimes I_{2N-1}$. Putting together the equalities, we get
\begin{align}
    I_{2N}\ast C = (I_{2N-1}\ast I_{2N}\ast C)\otimes \left(\frac{I_{2N-1}}{d_{2N-1}}\right),
\end{align}
which is equivalent to $_{2N}C= _{2N,2N-1}C$.  Repeating this process for $C_{N-1}$ and further for every $C_{i}, 1\le i\le N-1$, we end up with the recursive constraints in Eq.~(\ref{comb-def}). The positivity follows from the definition of Choi operator. 
 
Now we prove that any positive semidefinite operator $C$ satisfying Eq.~(\ref{comb-def}) is the Choi operator of a quantum comb. Define $C_i:=(\frac{I_{2i+1}}{d_{2i+1}}\ast  I_{2i+2}\ast\cdots\ast\frac{I_{2N-1}}{d_{2N-1}}\ast  I_{2N})\ast C$ to be the comb resulting from neutralizing all input-output pairs after the $i$-th step. From Eq.~(\ref{comb-def}) we have
\begin{align}
\Tr_{2,4,...,2i} C_i=I_{1,3,...,2i-1}
\end{align}
for all $1\le i\le N$, so in fact every $C_{i}$ is the Choi operator of a quantum channel $\map{C}_i$. With this, we use the Stinespring dilations of $\map{C}_i$ to construct a quantum circuit corresponding to the comb $C$. For $C_1$, there exists an isometry $W^{(1)}=\sum_{i}^{}|i\>_A \otimes K_i^{(1)}$, where $K_i^{(1)}$ is the Kraus operator of $\map{C}_1$ and $\left\{ |i\>_A \right\}$ are orthonormal states on an ancillary space $\h{A}$. Suppose that the isometry $W^{(N)}=\sum_{i}^{}|i\>_A \otimes K_i^{(N)}$ for $\map{C_N}$ with canonical Kraus operators $K_i^{(N)}$ is obtained by composing $N$ isometries, we need to show that the isometry $W^{(N+1)}=\sum_{i}^{}|i\>_B \otimes K_i^{(N+1)}$ is obtained by composing $N+1$ isometries. To this end, notice that for any state $\rho\in\otimes_{i=1}^{N}\h{2i+1}$, $\Tr_{2N+2}[C_{N+1}]=I_{2N+1} \otimes C_{N}$ implies
\begin{align}
\Tr_{2N+2} [\map{C}_{N+1}(\rho)]=\map{C}_{N} (\Tr_{2N+1} [\rho])
\end{align}
Therefore we have the same channel with two sets of equivalent Kraus operators $\left\{ \<m|K_i^{(N+1)} \right\}$ and 

$\left\{ K_j^{(N)} \otimes \<n| \right\}$, where $|m\>\in\h{2N+2}$, $|n\>\in\h{2N+1}$ and $K_i^{(k)}:\h{2k-1}\to\h{2k}$. Notice that $\left\{ K_j^{(N)} \otimes \<n| \right\}$ is canonical since $\left\{ K_j^{(N)} \right\}$ is canonical by assumption. Using the fact that equivalent Kraus operator representations are connected to the canonical one by an isometry, we can write
\begin{align}
\<m|K_i^{(N+1)}=\sum_{nj}^{} V_{mi,nj} K_j^{(N)} \otimes \<n|
\end{align}
which is equivalent to 
\begin{align}\label{K_i^{N+1}}
K_i^{(N+1)}=\<i|_B( I_{1,...,2N} \otimes V)( W^{(N)} \otimes I_{2N+1})
\end{align}
where $V:\h{2N+1}\otimes\h{A}\to\h{2N+2}\otimes\h{B}$ is an isometry defined by $V=\sum_{mi,nj}^{} V_{mi,nj} |m\>\<n| \otimes |i\>_{B}\<j|_{A}$. Inserting Eq.~(\ref{K_i^{N+1}}) into $W^{(N+1)}$, we have
\begin{align}
W^{(N+1)}=(I_{1,...,2N} \otimes V)(W^{(N)} \otimes I_{2N+1})
\end{align}
The map $\map{C}_{N+1}$ can then be written as
\begin{align}
\map{C}_{N+1} (\rho)=\Tr_{B} [(I \otimes V)(W^{(N)} \otimes I) \rho (W^{(N)\dagger} \otimes I)(I \otimes V^{\dagger})]
\end{align}
which is clearly a composition of $N+1$ isometries since $W^{(N)}$ is a composition of $N$ isometries by assumption. Thus the proof is completed. As a conclusion, Eq.~(\ref{comb-def}) is the necessary and sufficient condition for quantum comb.

\subsection{Circuit decomposition of quantum combs} \label{sec:circuit decomposition theory}
Given a generic $N$-step process $\map{C}$ with inputs and outputs $(\h{1},\h{2}),\dots,(\h{2N-1},\h{2N})$, is there always a quantum circuit that implements it? In this subsection, we provide a positive answer to this question in two steps. 

\threesubsection{Decomposition of a comb into isometries} First, we decompose $\map{C}$ (with Choi operator $C$) into $N$ isometries.  
By Ref.~\cite[Theorems 1 and 2]{Bisio2011PRA}, there exist isometries $\{V^{(k)}\}_{k=1}^N$ such that
\begin{equation}
    \map{C}(\rho) = \Tr_{A_N} \left[\left(V^{(N)}\otimes I_{2,4,\dots,2N-2}\right)\cdots \left(V^{(1)}\otimes I_{3,5,\dots,2N-1}\right)\rho \left(V^{(1)}\otimes I_{3,5,\dots,2N-1}\right)^\dagger \cdots \left(V^{(N)}\otimes I_{2,4,\dots,2N-2}\right)^\dagger \right]
\end{equation}
for any input state $\rho\in \map L\left(\otimes_{i=1}^N \map H_{2i-1}\right)$. Each isometry $V^{(k)}\in\map L\left(\map H_{2k-1} \otimes \map H_{A_{k-1}}, \map H_{2k} \otimes \map H_{A_{k}}\right)$ with minimal ancilla space is given by
\begin{equation}
    V^{(k)} = I_{2k} \otimes \left(C^{(k)*}\right)^{\frac 12}\left[|I\kk_{2k,(2k)'}I_{2k-1\rightarrow(2k-1)'}\otimes\left(C^{(k-1)*}\right)^{-\frac 12}\right],
\end{equation}
where $\map H_{A_{k}}=\Supp(C^{(k)*})$ is an ancillary space given by the support of $C^{(k)*}$ with $C^{(N)}$ defined in Eq.~(\ref{comb-def}) and $I_{2k+1}\otimes C^{(k)}=\Tr_{2k+2}C^{(k+1)}$, $\map H_{i'}$ is a copy of the Hilbert space $\map H_i$, $I_{2k-1\rightarrow(2k-1)'}:=\sum_i|i\>_{(2k-1)'}\<i|_{2k-1}$ is an identity map from $\map H_{2k-1}$ to $\map H_{(2k-1)'}$, and $\left(C^{(k-1)*}\right)^{-\frac 12}$ denotes the Moore–Penrose pseudoinverse of $\left(C^{(k-1)*}\right)^{\frac 12}$ with its support on $\map H_{A_{k-1}}$.

From this explicit construction it follows that the minimal dimension of the ancilla space for implementing the sequential strategy $C$ is $\dim(\map H_{A_{N}})=\rank(C^{(N)})\le d^{2N}$, having assumed the system dimension $d_i=d,\ \forall i$ for simplicity. Ref.~\cite{Pollock18PRA} proposed an alternative implementation based on iterative Stinespring dilation, which nevertheless requires an ancilla space of dimension $\prod_{j=0}^{N-1} d^{2(3^{j})}=d^{3^N-1}$. This approach of iterative Stinespring dilation is not optimal, because at each step it introduces a unitary instead of isometry to simulate the action of the comb on the system and all the ancillae introduced by previous steps so far, which is sometimes unnecessary.

\threesubsection{Decomposition of an isometry into universal quantum gates}

Second, any isometry\footnote{The isometry considered here is a $2^n \times 2^m$ matrix for integers $m\le n$; otherwise, we extend the isometry such that it has the desired dimension.} can be operationally implemented by a quantum circuit consisting of single-qubit gates and CNOT gates \cite{Iten2016PRA,Iten19Universal}. In practice, CNOT gates are usually much more costly than single-qubit gates, so it is desirable to achieve a CNOT count as low as possible. For the decomposition of a generic isometry from $m$ qubits to $n$ qubits with fewest CNOT gates, different methods need to be chosen for different $m$ and $n$ (see Ref.~\cite{Iten2016PRA} for reference). For the decomposition of a specific isometry, it is usually necessary to try out different methods in practice. Here we briefly introduce the main idea behind one of these decomposition methods, namely the column-by-column decomposition, which usually performs well for small $m$ and $n$, and in fact have the state-of-the-art performance in our example (see Subsection \ref{sec:optimal protocols comb decomposition}).

An isometry $V$ from $m$ to $n$ qubits ($m\le n$) can be represented in the matrix form by $V=U^{\dagger} I(2^n\times 2^m)$, where $U^{\dagger}$ is a $2^n\times 2^n$ unitary matrix and $I(2^n\times 2^m)$ is the first $2^m$ columns of the $2^n \times 2^n$ identity matrix. If we can obtain a decomposition of $U^\dagger$, then we simply need to initialize the state of the first $n-m$ qubits as $|0\>$ for implementing $V$. Equivalently, we can find a decomposition of $U$ such that $UV=I(2^n\times 2^m)$ and then inverse the circuit representing $U$. The idea is to find a sequence of unitary operations such that $U=U_{2^m-1}\cdots U_0$ transforms $V$ into $I(2^n\times 2^m)$ column by column. More specifically, we first choose a proper $U_0$ to map the first column of $V$ to the first column of $I(2^n\times 2^m)$, i.e., $U_0 V|0\>^{\otimes m}=|0\>^{\otimes n}$, then choose $U_1$ satisfying $U_1U_0 V|1\>^{\otimes m}=|1\>^{\otimes n}$ as well as $U_1U_0 V|0\>^{\otimes m}=|0\>^{\otimes n}$, \dots, until we determine $U_{2^m-1}$. The detailed algorithm can be found in Ref.~\cite{Iten2016PRA}, with the Mathematica implementation given by Ref.~\cite{Iten19Universal}.

\threesubsection{Circuit complexity of the decomposition}
For simplicity suppose $d_i=2$ for all $i=1,\dots,2N$. The number of CNOT gates for decomposing a general $N$-step quantum comb is exponential with respect to $N$. This can be seen by considering the two-step decomposition. 

In the first step, we decompose the comb into $N$ isometries. By construction, each isometry $V^{(k)}$ is a $d_{2k}\dim(\map H_{A_{k}}) \times d_{2k-1}\dim(\map H_{A_{k-1}})$ matrix. For the isometric implementation of an arbitrary comb $C^{(k)}$, we require $\dim(\map H_{A_{k}})= \max_{C^{(k)}: k\text{-step comb}}\rank(C^{(k)}) = (d_1d_2)^k=2^{2k}$. Therefore, each isometry $V^{(k)}$ maps a $(2k-1)$-qubit state to a $(2k+1)$-qubit state, having assumed $d_{2k}=d_{2k-1}=2$. 

In the second step, we decompose each isometry into single-qubit gates and CNOT gates. Decomposing an isometry from $m$ to $n$ qubits in general requires $2^{m+n}$ CNOT gates to leading order~\cite{Iten2016PRA}. Hence, decomposing the sequence $\{V^{(k)}\}_{k=1}^N$ requires $\map O(4^N)$ CNOT gates in total.

\section{Quantum strategy sets} \label{sec:strategy sets}
\subsection{Parametrized comb framework for quantum metrology} \label{sec:parametrized comb framework for quantum metrology}
Armed with quantum combs, we can now describe both the process to estimate and the strategy by their Choi operators.
Consider a generic quantum metrology task $(\map{C}_\phi,\strat)$.
The parametrized processes to estimate are described by a collection of $\phi$-dependent operators $\{C_\phi\}$, with each 
$$
C_\phi\in\text{Comb}[ ( \h{1},\h{2}),  \ldots, (\h{2N-1},\h{2N})],
$$
the set of quantum combs with input-output pairs $(\h{1},\h{2}),\dots,(\h{2N-1},\h{2N})$ as defined by Eq.~(\ref{comb-def}).

A strategy is described by a comb $P \in\strat$, the set of allowed strategies, that ``eats'' the parametrized process and ``spits'' a parametrized quantum state. For instance, when sequential strategies are allowed, the set of allowed strategies is $\seq = \text{Comb}[ (\mathbb C, \h{1}), (\h{2},\h{3}), \ldots, (\h{2N-2},\h{2N-1}), (\h{2N},\h{F})]$, with $\mathbb C$ denoting a trivial one-dimensional input space. 
It is guaranteed that the composition $P\ast C_\phi$ corresponds to a quantum state in $\L(\h{F})$, where $\h{F}$ is called the \emph{global future} space.

\begin{figure} [!htbp]
\centering
\subfigure[A sequential strategy.]{\includegraphics[width=0.5\textwidth]{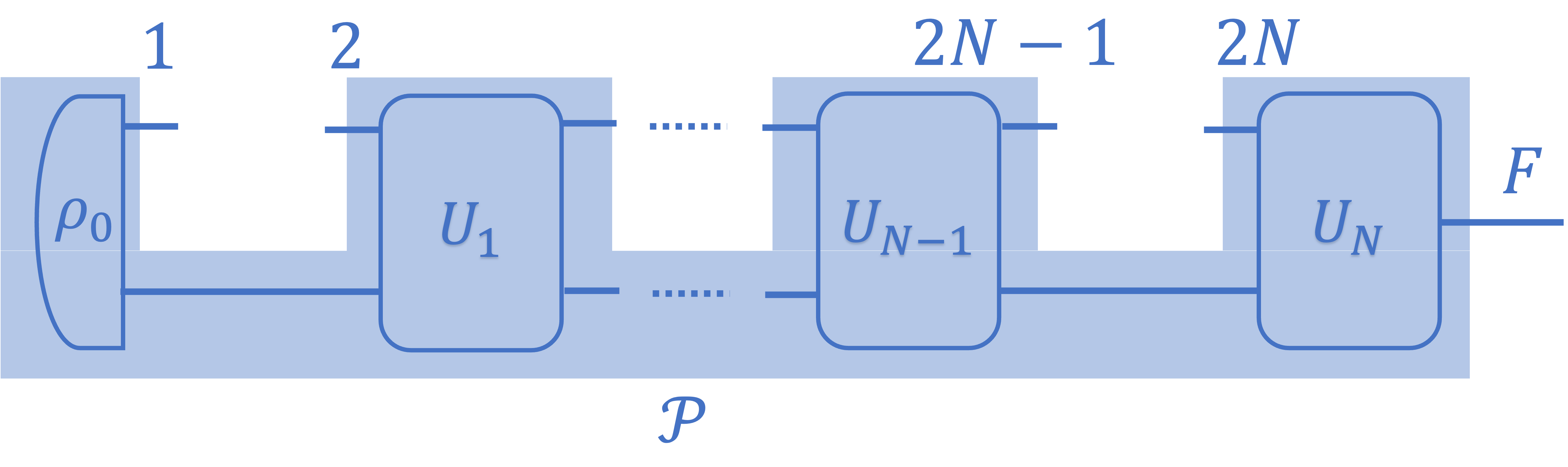}\label{fig:seq}} \quad
\subfigure[A parallel strategy.]{\includegraphics[width=0.25\textwidth]{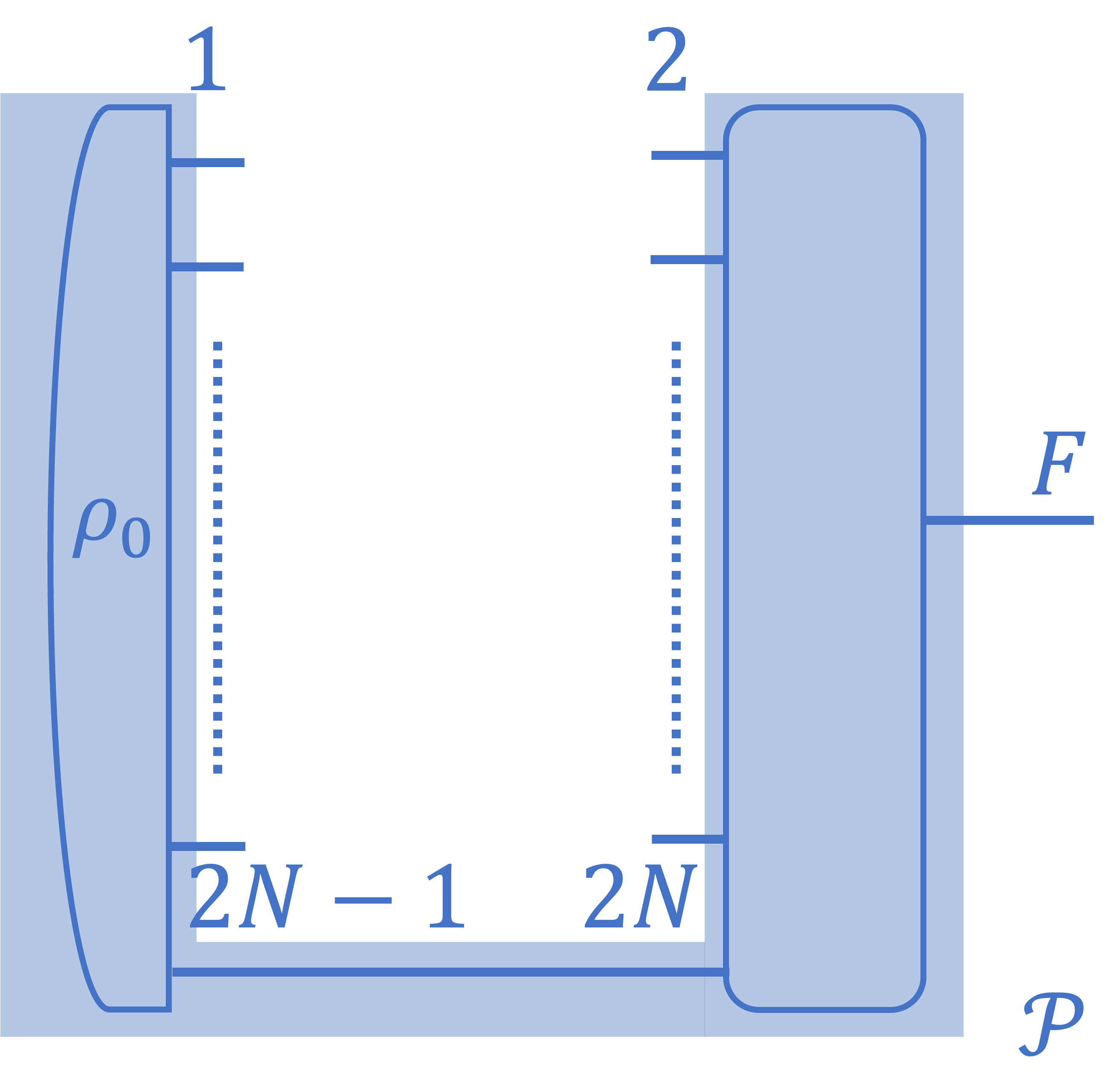}\label{fig:par}}
\caption{Strategies with definite causal orders.}
\end{figure}

In practice, the set of allowed strategies is determined by both the physical constraints and the goal of the experiment. The set of all strategies under consideration, $\strat$, is a subset of positive semidefinite operators on $\h{1}\otimes\cdots\otimes\h{2N}\otimes\h{F}$.

It is important to note that, in this tutorial, we will consider \emph{pure} strategies with the requirement $\rank(P)=1$. This is naturally motivated by the monotonicity of QFI (cf.~Corollary~\ref{cor_monotone_qfi}), and discarding any part of the probe state cannot increase the obtained information.
In addition, the requirement also helps to obtain the exact value of the maximal QFI.
Nevertheless, some practical constraints (for instance, if the device has noise or limited memory) do not fulfil this requirement. This will be addressed in future works.

\subsection{Strategies with definite causal orders}\label{subsec_defco_strategy}

\threesubsection{Parallel strategies (Par)}
The family of parallel strategies (see Figure~\ref{fig:par}) is the first and one of the most successful examples of quantum-enhanced metrology, featuring the usage of entanglement to achieve precision beyond the classical limit \cite{Giovannetti2006PRL}. A parallel strategy set $\mathsf{Par}$ is defined as the collection of $P \in \L(\h{1}\otimes\cdots\otimes\h{2N}\otimes\h{F})$ such that \cite{Chiribella2009PRA}
\begin{equation} \label{eq:parallel strategy set}
    \begin{aligned}
        &P\ge0,\ \rank(P)=1,\ \Tr P=\prod_{i=1}^N d_{2i},\\ 
        &_F{P}= _{F,2,4,\dots,2N}{P}.
    \end{aligned}
\end{equation}

\threesubsection{Sequential strategies (Seq)}
A more general protocol is to allow for sequential use of $N$ channels assisted by ancillae, where the input of the latter channel cannot affect the output of the former channel, and any control gates can be inserted between channels (see Figure~\ref{fig:seq}). A sequential strategy set $\mathsf{Seq}$ is defined as the collection of $P \in \L(\h{1}\otimes\cdots\otimes\h{2N}\otimes\h{F})$ such that \cite{Chiribella2009PRA}
\begin{equation} \label{eq:sequential strategy set}
    \begin{aligned}
    & P\ge0,\ \rank(P)=1,\ \Tr P=\prod_{i=1}^N d_{2i},\ _{F} P= _{F,2N} P,\\
    &_{F,2N,\dots,2i} P= _{F,2N,\dots,2i+1} P\quad i=1,\dots, N-1.
    \end{aligned}
\end{equation}
Unlike the case of parallel strategies, there is no existing way of evaluating the exact QFI using sequential strategies. 

\subsection{Strategies with quantum superpositions of causal orders}\label{subsec_sup_strategy}

Recently, new spacetime structures have been conjectured, such that quantum theory is valid locally but no global causal order is predetermined \cite{Chiribella2013PRA,Oreshkov2012}. For example, the global spacetime may admit superpositions of causal orders, which allows for the following two families of strategies.

\begin{figure} [!htbp] 
    \centering
\subfigure[A quantum SWITCH strategy.]{\includegraphics[width=0.4\textwidth]{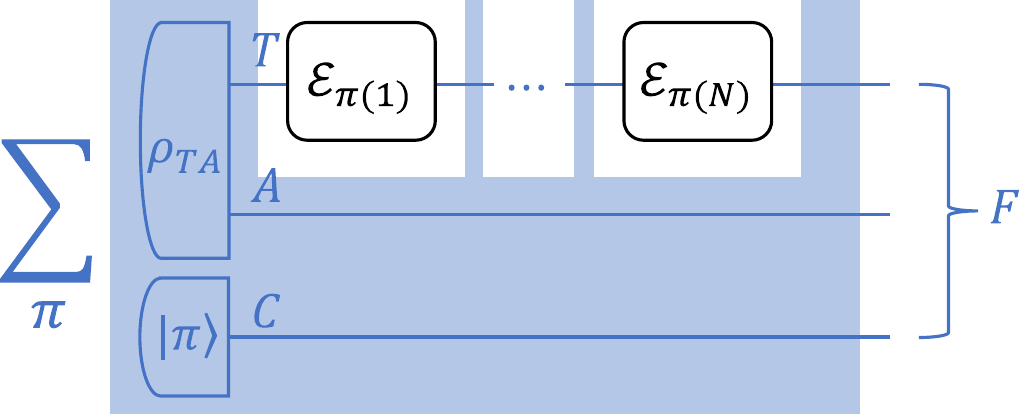}\label{fig:swi}} \quad
\subfigure[A causal superposition strategy.]{\includegraphics[width=0.55\textwidth]{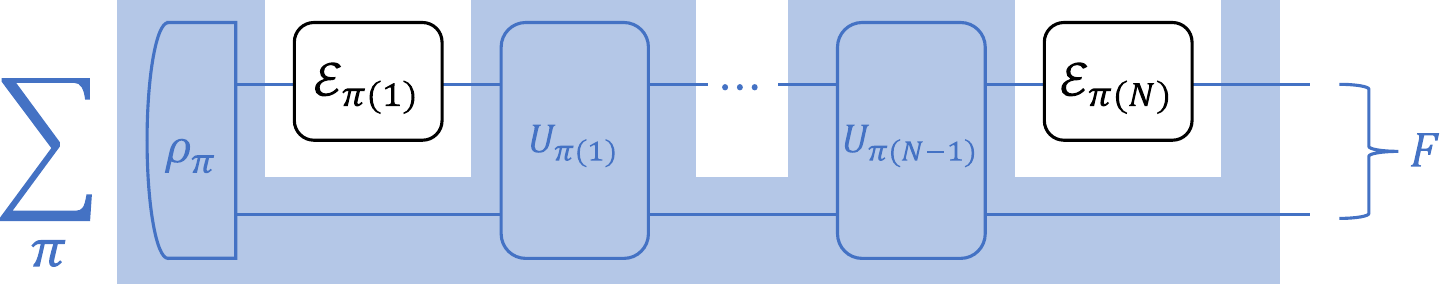}\label{fig:sup}}
\caption{Strategies with superposition of causal orders. Note that the ``summation'' over permutation $\pi$ should not be understood as a probabilistic mixture of different causal orders, as shown in Eq.~(\ref{eq:SWITCH Kraus operators}). For the quantum SWITCH strategy, the figure illustrates the use a separable input state, while in general one can use an arbitrary state $\rho_{TAC}$.}
\end{figure}

\threesubsection{Quantum SWITCH strategies (SWI)}
The first one, denoted by $\mathsf{SWI}$, takes advantage of the (generalized) quantum SWITCH \cite{COLNAGHI20122940,Araujo2014PRL}, where the execution order of $N$ channels is entangled with the state of an $N!$-dimensional control system  [see Figure~\ref{fig:swi}]. 
A quantum SWITCH strategy set $\mathsf{SWI}$ is defined as the collection of $P \in \L(\h{1}\otimes\cdots\otimes\h{2N}\otimes\h{F})$ such that
\begin{equation}\label{SWI_constraints}
    \begin{aligned}
        &\rank(P) = 1,\ P\ge0,\\
        &P = \left(\rho_{T,A,C} \right) * \dyad{P^{(\mathrm{SW})}},\ \rho_{T,A,C} \ge 0,\ \Tr \rho_{T,A,C} = 1,
    \end{aligned}
\end{equation}
where $|P^{(\mathrm{SW})}\>
:= \lvert I \kk_{A,F_A}\sum_{\pi\in S_N}\left[|\pi\>_C \lvert I \kk_{T,2\pi(1)-1}\left(\otimes_{i=1}^{N-1}\lvert I \kk_{2\pi(i),2\pi(i+1)-1}\right) \lvert I \kk_{2\pi(N),F_T} |\pi\>_{F_C}\right]$ corresponds to a (generalized) quantum SWITCH for $N$ operations, each permutation $\pi$ is an element of the symmetric group $S_N$ whose order is $N!$, and $\{|\pi\>_C\}$ forms an orthonormal basis. $\rho_{T,A,C}$ is the joint state of the control of the SWITCH, the target system, and the ancilla. We assume each $\map H_i$ for $i=1,\dots,2N$ has the same dimension $d_1$. $\map H_T\simeq \map H_i$ denotes the input space of the target system, $\map H_A$ the ancillary space, and $\map H_C$ the space of the control system. Correspondingly, $\map H_{F_T}$, $\map H_{F_A}$ and $\map H_{F_C}$ denote the future output spaces of each part. The global future space $\map H_F = \map H_{F_T} \otimes \map H_{F_A} \otimes \map H_{F_C}$.

To understand the constraints for the quantum SWITCH, we leave the ancilla aside and see that the process conditioned on the $k$-th causal order is 
\begin{align}
\Tr_{F_C} [|k\>\<k|_{F_C} \dyad{P^{(\mathrm{SW})}}]=|k\>\<k|_{C} \otimes |I\kk \bb I|_{T,2\pi(i)-1 }  \left(\otimes_{i=1}^{N-1}\lvert I \kk_{2\pi(i),2\pi(i+1)-1}\right) \otimes |I\kk \bb I|_{2\pi(N),F_T},  
\end{align}
where $|k\>\<k|_{C}$ is the initial state of the control system and the right hand side is the process matrix of the composed identity maps conditioned on $|k\>\<k|_{C}$. This means that while the quantum SWITCH generates coherent superposition of different causal orders of the channels, it allows no intermediate controls, but only identity maps within each causal order. Thus in the circuit there are only straight lines in the internal structure of each comb representing one causal order in the quantum SWITCH.

We remark that even the quantum SWITCH of identical channels may have a nontrivial effect \cite{Ebler2018PRL}. Consider two channels described by Kraus operators $\{K^{(1)}_i\}_i$ and $\{K^{(2)}_i\}_i$, respectively. By inserting these two channels into a quantum SWITCH, the output channel is described by Kraus operators
\begin{equation} \label{eq:SWITCH Kraus operators}
    K^{(2)}_iK^{(1)}_j \otimes \dyad{0}_C + K^{(1)}_jK^{(2)}_i \otimes \dyad{1}_C.
\end{equation}
Even if $\{K^{(1)}_i\}_i$ and $\{K^{(2)}_i\}_i$ are identical, the quantum SWITCH may create some correlations between the target system and the control system, as shown in Ref.~\cite{Ebler2018PRL}.

\threesubsection{Causal superposition strategies (Sup)}
More generally, we consider the quantum superposition of multiple sequential orders, each with a unique order of querying the $N$ channels [see Figure~\ref{fig:sup}]. This can be implemented by entangling $N!$ definite causal orders with a quantum control system \cite{Wechs2021PRXQuantum}. If $N=2$ and the control system is traced out, this notion is equivalent to causal separability \cite{Oreshkov2012,Araujo2015}. A causal superposition strategy set $\mathsf{Sup}$ is defined as the collection of $P \in \L(\h{1}\otimes\cdots\otimes\h{2N}\otimes\h{F})$ such that
\begin{equation}\label{SUP_constraints}
    \begin{aligned}
    &\rank(P) = 1,\ P\ge0\\
    &\Tr_FP= \sum_\pi q^\pi P^{\pi},\ \sum_{\pi\in S_N} q^\pi=1, \\
    &P^{\pi} \in \mathsf{Seq}^{\pi},\ q^\pi \ge 0,\ \pi\in S_N, 
    \end{aligned}
\end{equation}
where each permutation $\pi$ is an element of the symmetric group $S_N$ of degree $N$, and each $\mathsf{Seq}^{\pi}$ denotes a sequential strategy set whose execution order of $N$ channels is $\map E_\phi^{\pi(1)}\rightarrow\map E_\phi^{\pi(2)}\rightarrow\cdots\rightarrow\map E_\phi^{\pi(N)}$, having denoted by $\map E_\phi^k$ the channel from $\map L(\map H_{2k-1})$ to $\map L(\map H_{2k})$. We remark that $\mathsf{SWI}$  is a subset of $\mathsf{Sup}$, where the intermediate control is trivial. 

Comparing Eq.~(\ref{SUP_constraints}) with the constraints for $\mathsf{SWI}$, we see that $\mathsf{Sup}$ is generalized from $\mathsf{SWI}$ by replacing the composed identity process conditioned on each causal order $\pi$ by some sequential strategy $P^{\pi} \in \mathsf{Seq}^{\pi}$ with probability $q^\pi$, meaning that instead of enforcing identity maps, now we allow general intermediate controls under each causal order.

\subsection{Strategies with general indefinite causal orders}\label{subsec_ico}

The previously introduced strategies, including those in $\mathsf{SWI}$ and $\mathsf{Sup}$, do not violate any causal inequality after tracing out the global future \cite{Oreshkov2012,Purves2021,Wechs2021PRXQuantum}, as they become classical mixtures of fixed-causal-order strategies. More explicitly, a causal witness is a Hermitian operator $W$, such that 
\begin{align}
    \Tr[WC]\ge 0,
\end{align}
for any causally separable process matrix $C$. $\mathsf{SWI}$ and $\mathsf{Sup}$ contain causally non-separable strategies when the global future is kept coherently, i.e., for $C\in\mathsf{SWI}\subset\mathsf{Sup}$, there exists a Hermitian $W$ such that $\Tr[WC]<0$. However, tracing out the global future, there is no witness $\tilde{W}$ such that $\Tr[\tilde{W}\tilde{C}]<0$, where $\tilde{C}:=\Tr_F[C]$ for $C\in\mathsf{SWI}\subset\mathsf{Sup}$. 

Here we introduce the most general family of indefinite-causal-order strategies, some of which do violate certain causal inequalities even after tracing out the global future\footnote{Note that some strategies in $\mathsf{ICO}$, such as quantum circuits with quantum controlled casual order (QC-QCs) \cite{Wechs2021PRXQuantum,Purves2021}, are not in $\mathsf{Sup}$ but also do not violate any causal inequality.}. Here the only requirement is that the concatenation of the strategy $P$ with $N$ arbitrary channels with extended ancilla space results in a legitimate quantum channel. The causal relations in this case  \cite{Araujo2015} are a bit cumbersome, but for our purpose what matters is the dual affine space (see Theorem \ref{thm_qfi_sdp}), which is simply the space of no-signaling channels \cite{Chiribella2013PRA,Chiribella_2016} (see Subsection~\ref{sec:dual affine spaces}). 

A general indefinite-causal-order strategy set, denoted as $\mathsf{ICO}$, is defined as the collection of $P \in \L(\h{1}\otimes\cdots\otimes\h{2N}\otimes\h{F})$ such that
\begin{equation}\label{def-general-ico}
    \begin{aligned}
        &\rank(P) = 1,\ P\ge0,\\ 
        &P*\left(\otimes_{j=1}^N E^j\right) \ge 0,\ \Tr_{F,A_2,A_4,\dots,A_{2N}} \left[P*\left(\otimes_{j=1}^N E^j\right)\right] = I_{A_1,A_3,\dots,A_{2N-1}}, 
    \end{aligned}
\end{equation}
for any $E^j \in \map L(\map H_{2j-1}\otimes\map H_{2j}\otimes \map H_{A_{2j-1}} \otimes \map H_{A_{2j}})$ that denotes the Choi operator of an arbitrary quantum channel $\map{E}_j$ with an arbitrary ancillary input space $\map H_{A_{2j-1}}$ and output space $\map H_{A_{2j}}$. The physical meaning of Eq.~(\ref{def-general-ico}) is that the composition of an ICO strategy and arbitrary $\{\map{E}_j\}$ results in a quantum channel from $\otimes_{j=1}^{N}\spc{H}_{A_{2j-1}}$ to $(\otimes_{j=1}^{N}\spc{H}_{A_{2j}})\otimes\spc{F}$.

For $N=2$, $\mathsf{ICO}$ is defined by the following constraints:
\begin{align}
    &\rank(P) = 1,\ P\ge0,\ \Tr P = d_2\cdot d_4,\\
    &\ _{F,1,2} P = \ _{F,1,2,4} P,\\
    &\ _{F,3,4} P = \ _{F,2,3,4} P,\\
    & _{F} P = _{F,4} P + _{F,2} P - _{F,2,4} P,
\end{align}
which are equivalent to the conditions in Eq.~(\ref{def-general-ico}) (see \cite[Appendix B]{Araujo2015} for the proof). A concrete example is the Oreshkov-Costa-Brukner (OCB) process \cite{Oreshkov2012}, defined as any purification\footnote{Here a purification is defined as an operator, of which by tracing out the global future we obtain the original operator. The purification of an ICO strategy is still an ICO strategy. Note that in Ref.~\cite{Araujo2017purification} a pure process is defined as a supermap which preserves the unitarity, and the purification in that context has a different operational meaning, such that some ICO processes are not ``purifiable'' therein.} $P_{OCB}\in\spc{L}(\h{F}\otimes\h{1}\otimes\h{2}\otimes\h{3}\otimes\h{4})$ of the following process matrix:
\begin{align}
    _F P_{OCB} = \frac{1}{4}\left[I_{1,2,3,4}+\frac{I_1\otimes Z_{2}\otimes Z_{3}\otimes I_4+Z_{1}\otimes I_2\otimes X_{3}\otimes Z_{4}}{\sqrt{2}}\right],
\end{align} 
where $Z(X)$ denotes the Pauli-$Z(X)$ operator. A causal witness for $P_{OCB}$ is 
\begin{align}
    W_{OCB} = \frac{1}{4}\left(I_{1,2,3,4}-I_1\otimes Z_{2}\otimes Z_{3}\otimes I_4-Z_{1}\otimes I_2\otimes X_{3}\otimes Z_{4}\right)\otimes I_F.
\end{align}

We note that, unlike the previous strategies that can always be physically realized, the physical realization of strategies in $\mathsf{ICO}$ (including $P_{OCB}$) is untraceable \cite{Purves2021,Wechs2021PRXQuantum}. The optimal value obtained with general $\mathsf{ICO}$ nevertheless serves as a useful tool that can gauge the performances of different strategies. For example, as we will show, in some cases the optimal QFI $J^{(\mathsf{Sup})}$ and $J^{(\mathsf{ICO})}$ are equal or nearly equal. This then shows that the physically realizable strategy obtained from the set $\mathsf{Sup}$ is already optimal or nearly optimal among all possible strategies, which we will not be able to tell without $J^{(\mathsf{ICO})}$.

\begin{figure} [!htbp]
\centering
\subfigure[Strategy sets.]{\includegraphics[width=0.45\textwidth]{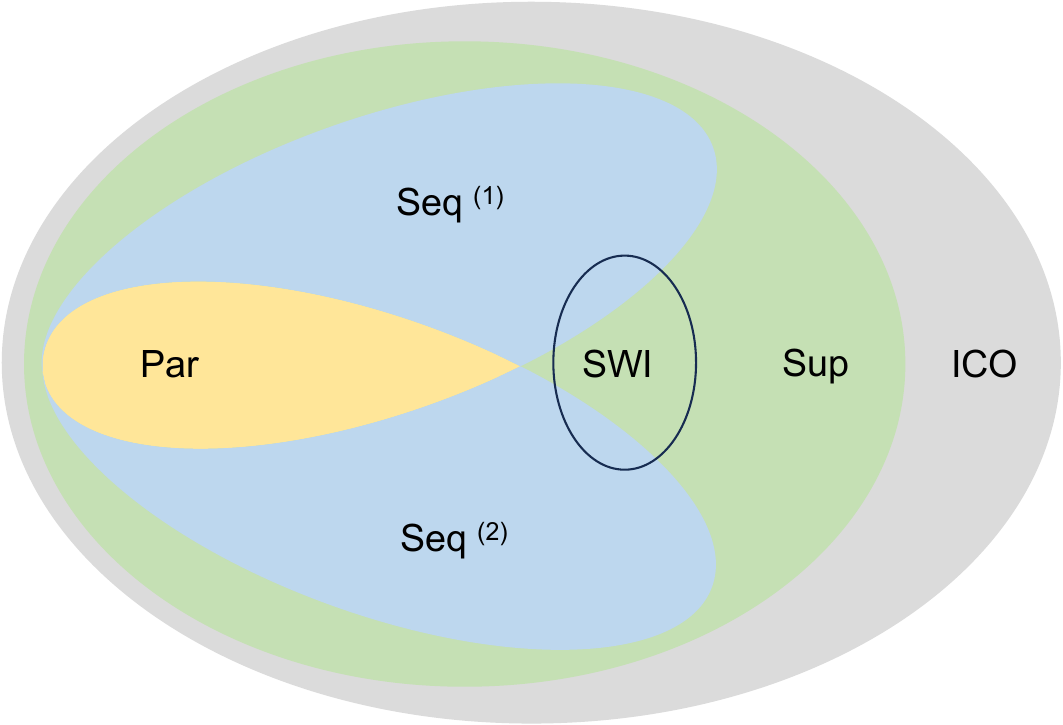}\label{subfig:relations_strategies}} \quad
\subfigure[Dual affine spaces.]{\includegraphics[width=0.45\textwidth]{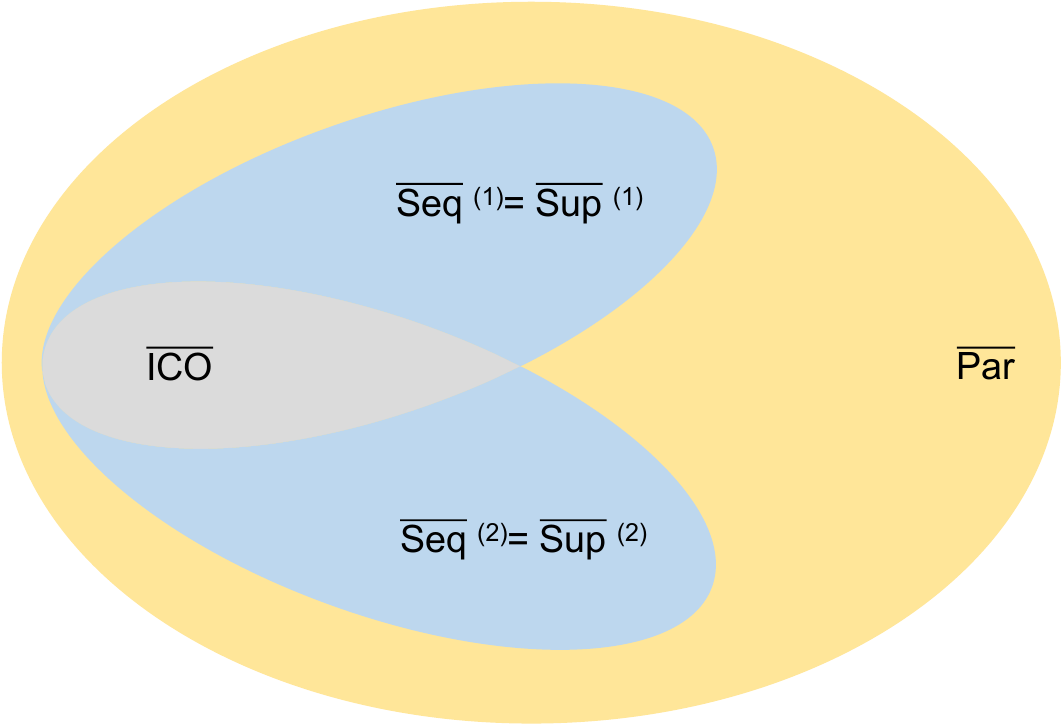}\label{subfig:relations_dual_strategies}}
\caption{\label{fig:relations}Relations between strategies sets and the corresponding dual affine spaces for estimating $N=2$ channels. $\mathsf{Seq}^{(1)}$ and $\mathsf{Seq}^{(2)}$ correspond to two different sequential orders.}
\end{figure}

The inclusion relations between different strategy sets for $N=2$ are schematically illustrated in Figure~\ref{fig:relations}\protect\subref{subfig:relations_strategies}.

\section{Optimal quantum metrology under strategy constraints} \label{sec:optimal metrology}
In this section, we derive the main result---an approach to the optimal QFI and the optimal strategies.
\subsection{The task QFI}

The core question is what is the ultimate precision limit given the process to estimate and the set of allowed strategies, captured by the task QFI \cite{Yang2019PRL}:
\begin{defi}[The task QFI]
    For a metrology task $(\map{C}_\phi,\strat)$ [cf.~Def.~\ref{defi-metro-task}], its task QFI\footnote{We assume that the rank of the output state $P\ast C_\phi$ is independent of $\phi$, to avoid the subtle issue of the discontinuity of QFI at the rank changing point. See Refs.~\cite{Dominik17PRA,Seveso_2020,zhou2019exact,Ye22PRA} for related discussions.} is defined as: 
  \begin{equation} \label{def_comb_qfi}
    J(\map{C}_\phi,\strat) := \max_{P\in \strat} J(P\ast C_\phi),
  \end{equation}
  where, on the right hand side (R.H.S.), $J(\cdot)$ denotes the QFI of a quantum state $(\cdot)$ [cf.~Eq.~(\ref{def_state_qfi})], and $C_\phi$ denotes the Choi operator of $\map{C}_\phi$.   
\end{defi} 
  The QFI gives a lower bound on the variance of any unbiased estimator $\hat{\phi}$ via the quantum Cram\'{e}r-Rao bound \cite{helstrom1976quantum,holevo2011probabilistic}, which can be extended to the generic setting of metrology tasks \cite{Yang2019PRL}:
  By the quantum Cram\'{e}r-Rao bound \cite{helstrom1976quantum,holevo2011probabilistic}, the mean squared error of estimating $\phi$ fixing a strategy $P$ is bounded as $\text{MSE}(\phi,P)\ge 1/(\nu J(P\ast C_\phi))$, where $\nu$ is the number of times that the experiment is repeated. Optimizing the strategy within the allowed strategy set and using Eq.~(\ref{def_comb_qfi}) yield
    \begin{equation}\label{comb-crb}
    \text{MSE}(\phi,\strat):=\min_{P\in\strat}\text{MSE}(\phi,P)\ge \frac{1}{\nu J(\map{C}_\phi,\strat)}.
    \end{equation}
    Since here $\phi$ is a single-parameter the bound is tight (in the limit of $\nu\to\infty$). It is thus established that the task QFI $J(\map{C}_\phi)$, as defined in Eq.~(\ref{def_comb_qfi}), amounts to the ultimate precision of the task.

\subsection{Overview of the main results}

Here we provide the main results, including a semi-definite program (SDP) for the maximal QFI and an algorithm for the optimal strategy, and delegate their proofs to later subsections.

\threesubsection{An SDP for the task QFI}
Consider a metrology task $(\map{C}_\phi,\strat)$. Denoting by $r:=\max_\phi\rank(C_\phi)$, since $C_\phi\ge0$, we can find a set of vectors $\{|C_{\phi,i}\>\}$ such that
\begin{align}\label{C-decomp}
C_\phi = \sum_{i=1}^{r}|C_{\phi,i}\>\<C_{\phi,i}| = \mathbf{C}_\phi\mathbf{C}_\phi^\dag,
\end{align}
where $\mathbf C_\phi:=(|C_{\phi,1}\>,\dots,|C_{\phi,r}\>)$.
\footnote{We assume $\left\{|C_{\phi,i}\>\right\}$ is continuously differentiable with respect to $\phi$.} 
Note that the decomposition is not unique, but it is sufficient to find an arbitrary decomposition. The QFI can be obtained by performing optimization over all possible decompositions, taking an $r$-dimensional Hermitian matrix $h$ as the variable. In addition, to obtain a simple expression of the SDP, we need to put some constraints on the form of the strategy set $\strat$. 
Explicitly, we require that there exist affine spaces of Hermitian operators $\{\mathsf S^i\}$ for $i=1,\dots,K$ such that:
\begin{align}\label{req_strategy_set}
\strat&=\left\{P\in\L(\h{1}\otimes\cdots\otimes\h{2N}\otimes\h{F})\mid~P\ge0,~\rank(P)=1,\Tr_F[P]\in\widetilde{\strat}\right\},\\
\widetilde{\strat} & := \Conv\left\{\bigcup_{i=1}^{K}\left\{S^i\ge0\middle|S^i\in\mathsf S^i\right\}\right\},
\end{align}
where $\Conv\{\cdot\}$ denotes the convex hull. It is noteworthy that all strategy sets ($\mathsf{Par},\mathsf{Seq},\mathsf{SWI},\mathsf{Sup},\mathsf{ICO}$) introduced in Sections \ref{subsec_defco_strategy}--\ref{subsec_ico} satisfy the above requirement.
\begin{theo}[An SDP for the task QFI]\label{thm_qfi_sdp}
Given a metrology task $(\map{C}_\phi,\strat)$ with $\strat$ satisfying Eq.~(\ref{req_strategy_set}), the task QFI equals the solution to the following SDP:
\begin{equation}
    \begin{aligned}
    J(\map{C}_\phi,\strat) =& \min_{\lambda,Q^i,h} \lambda, \\
    \mathrm{s.t.}\ &\,A^i \ge 0,\ Q^i\in \overline{\mathsf S}^i,\ i=1,\dots,K,
    \end{aligned} 
\end{equation}
having defined \begin{equation} \label{eq:Schur complement}
    A^i = \left( 
\begin{array}{c | c} 
  \frac \lambda 4 I\left(r\times r\right) & 
  \begin{array}{c}
       \<\overline{\dot{\tilde{C}}_{\phi,1}}| \\
       \vdots \\
       \<\overline{\dot{\tilde{C}}_{\phi,r}}|
  \end{array}\\ 
  \hline 
  \begin{array}{ccc}
       |\overline{\dot{\tilde{C}}_{\phi,1}}\> & \hdots & |\overline{\dot{\tilde{C}}_{\phi,r}}\>
  \end{array} & Q^i
 \end{array} 
\right),
\end{equation}
where $|\dot{\tilde{C}}_{\phi,i}\>=|\dot{C}_{\phi,i}\>-\mathrm i\sum_j|C_{\phi,j} \> h_{ji}$, $h$ is an $r\times r$ Hermitian matrix, $I(r \times r)$ is an $r\times r$ identity matrix, $\overline{(\cdot)}$ denotes the complex conjugate of $(\cdot)$, $\dot C_{\phi,i}$ denotes the derivative of $C_{\phi,i}$ with respect to $\phi$, and $\overline{\mathsf {S}}^i:=\left\{Q \mid Q^\dag=Q, \Tr(QS)=1,S\in\mathsf S^i\right\}$ is the dual affine space of $\mathsf S^i$.
\end{theo}

The proof can be found in Section \ref{subsection_qfi_sdp}. We remark that the framework applies to both Markovian tasks, where $C_\phi = E_\phi^{\otimes N}$ with $E_\phi$ being the Choi operator of some parametrized channel, and non-Markovian tasks. For non-Markovian tasks, $C_\phi$ is a non-product operator and the correlation captures the underlying memory of the process.

\threesubsection{An algorithm for optimal strategies}
We also design an algorithm that yields a strategy attaining the task QFI.
Assuming the strategy set to satisfy the requirement in Eq.~(\ref{req_strategy_set}), the algorithm runs as follows:\\
\begin{algorithm}[!htbp] 
\caption{Find an optimal strategy for a metrology task $(\map{C}_\phi,\strat)$.}\label{alg_optimal_strategy}
\begin{enumerate}
    \item Solve for an optimal value $h=h^{(\mathrm{opt})}$ of the SDP in Theorem \ref{thm_qfi_sdp}.
    \item By SDP find a solution $\tilde P^{(\mathrm{opt})}$ to the maximization problem 
    \begin{equation} 
    \max_{\tilde P\in\widetilde{\strat}} \Tr\left[\tilde{P}\Omega_\phi(h^{(\mathrm{opt})})\right],
\end{equation}
 such that
    \begin{align}\label{opt_strategy_condition} 
        \mathbf C_\phi^\dagger \tilde P^T \left(\dot{\mathbf C}_\phi-\mathrm i \mathbf C_\phi h^{(\mathrm{opt})}\right)
    \end{align}
    is Hermitian,
    where $\mathbf C_\phi := \left(|C_{\phi,1}\>,\dots,|C_{\phi,r}\>\right)$. 
    \item Purify $\tilde P^{(\mathrm{opt})}$ on a global future $\h{F}$ and output the resultant strategy $P^{(\mathrm{opt})} \in \strat$.
\end{enumerate}
\end{algorithm}
The validity of the algorithm is expounded in Section \ref{sec:optimal strategy}. The output of the above algorithm is in the form of a Choi operator, i.e., a matrix in $\L(\h{1}\otimes\cdots\otimes\h{2N}\otimes\h{F})$.
 For strategies following definite causal order, there exists an operational method of mapping the Choi operator of the strategy to a probe state and a sequence of in-between control operations with minimal memory space \cite{Bisio2011PRA}. For causal order superposition strategies (see the strategy set $\mathsf{Sup}$), they can always be implemented by controlling the order of operations in a circuit with a quantum SWITCH \cite{Liu23PRLoptimal}. 

\subsection{Proof of Theorem \ref{thm_qfi_sdp}: an SDP for the task QFI}\label{subsection_qfi_sdp}
By the basic property of purification, Eq.~(\ref{def_state_qfi}) can be cast into the following form:
\begin{equation}\label{qfi_state}
    J(\rho_\phi) = 4\min_{|\psi_{\phi,i}\>}\sum_{i=1}^q\Tr\left(\dyad{\dot{\psi}_{\phi,i}}\right),
\end{equation}
for any integer $q \ge \rank(\rho_\phi)$, where $\left\{|\psi_{\phi,i}\>\right\}$ is a set of unnormalized vectors such that $\rho_\phi = \sum_i\dyad{\psi_{\phi,i}}$\footnote{We assume that $\left\{|\psi_{\phi,i}\>\right\}$ is continuously differentiable with respect to $\phi$ and $\rho_\phi$ has a constant rank for all $\phi$.}. Combining with Eq.~(\ref{def_comb_qfi}) [and noticing the definition of the link product (\ref{def_link_product})], we have
\begin{align}
P\ast C_\phi = \sum_{i=1}^r(\<\overline{C_{\phi,i}}|\otimes I_F)|P\>\<P|(|\overline{C_{\phi,i}}\>\otimes I_F)
\end{align}
where $\overline{(\cdot)}$ denotes the complex conjugate of $(\cdot)$ and $\{|C_{\phi,i}\>\}$ are the vectors in the decomposition (\ref{C-decomp}). We have restricted the strategy to be pure $P=|P\>\<P|$ due to the monotonicity of QFI under quantum channels, as explained in Section \ref{sec:parametrized comb framework for quantum metrology}. Note that the decomposition is not unique, and any two decompositions are connected by a unitary:
\begin{align}\label{comb_decomp_u}
\mathbf C'_\phi=\mathbf C_\phi V_\phi.
\end{align}

Noticing that $\{(\<\overline{C_{\phi,i}}|\otimes I_F)|P\>\}$ are vectors on $\h{F}$ whose derivatives are $\{\<\overline{\dot{C}_{\phi,i}}|\otimes I_F)|P\>\}$, we can apply the state QFI formula (\ref{qfi_state}), and express the QFI of the signal process as
\begin{align}\label{qfi_inter1}
J^{(\strat)}(\map{C}_\phi)=\max_{P\in\strat}\min_{\mathbf{C}_\phi}\Tr\left[P(I_F\otimes\Omega(\mathbf{C}_\phi))\right],
\end{align}
where $\Omega(\mathbf{C}_\phi)$ is the performance operator defined as
\begin{align}
\Omega(\mathbf{C}_\phi) = 4\sum_i\left(|\dot{C}_{\phi,i}\>\<\dot{C}_{\phi,i}|\right)^T=4\left(\mathbf{\dot{C}}_\phi\mathbf{\dot{C}}_\phi^\dag\right)^T.
\end{align}
From Eq.~(\ref{comb_decomp_u}), the freedom of $\mathbf{C}_{\phi}$ is encoded in the unitary $V_\phi$. Denoting by $h:=i\dot{V}_\phi V_\phi^\dag$, we have, for an arbitrary $\tilde{\mathbf{C}}_\phi:=\mathbf{C}_\phi V_\phi$:
\begin{align}
\dot{\tilde{\mathbf{C}}}_\phi=(\dot{\mathbf{C}}_\phi-i\mathbf{C}_\phi h )V_\phi \quad h:=i\dot{V}_\phi V^\dag_\phi.
\end{align}
The freedom is then captured by a $r$-dimensional Hermitian matrix $h$. We can fix one decomposition and rewrite the performance operator as
\begin{align} \label{eq:performance_op}
\Omega_\phi(h) = 4\left((\dot{\mathbf{C}}_\phi-i\mathbf{C}_\phi h)(\dot{\mathbf{C}}_\phi-i\mathbf{C}_\phi h)^\dag\right)^T.
\end{align}
Eq.~(\ref{qfi_inter1}) becomes
\begin{align}
J^{(\strat)}(\map{C}_\phi)=\max_{\tilde{P}\in\widetilde{\strat}}\min_{h:h^\dag=h}\Tr\left[\tilde{P}\Omega_\phi(h)\right],
\end{align}
where $\widetilde{\strat}:=\{\tilde{P} = \Tr_F P~:~\exists\,P\in\strat\}$.

Next, we exchange the order of minimization and maximization thanks to Fan's minimax theorem \cite{Fan1953}, since the objective function is concave on $P$ and convex on $h$, and $\strat$ is a compact set: 
\begin{equation} \label{qfi_inter2}
    J^{(\strat)}(\map{C}_\phi)=\min_{h:h^\dag=h}\max_{\tilde{P}\in\widetilde{\strat}}\Tr\left[\tilde{P}\Omega_\phi(h)\right].
\end{equation}
Reformulating the condition of Theorem 1, we require that each operator $\tilde P\in\strat$ can be written as a convex combination of positive semidefinite operators $S^i$, $i=1,\dots,K$:
\begin{equation}
    \tilde P = \sum_{i=1}^Kq^iS^i,\ \mathrm{for}\ \sum_{i=1}^Kq^i=1,\ q^i\ge0,\ S^i\ge0,\ S^i\in\mathsf S^i,\ i=1,\dots,K,
\end{equation}
where each $\mathsf S^i$ is an affine space of Hermitian operators. Thus Eq.~(\ref{qfi_inter2}) can be reformulated as
\begin{equation}
    \begin{aligned}
    J^{(\strat)}(\map{C}_\phi) =& \min_h\max_{\tilde P}\Tr\left[\tilde P\Omega_\phi(h)\right], \\
    \mathrm{s.t.}\ &\,\tilde P=\sum_{i=1}^Kq^iS^i,\ \sum_{i=1}^Kq^i=1, \\
    &\,q^i\ge0,\ S^i\ge0,\ S^i\in\mathsf S^i,\ i=1,\dots,K.
    \end{aligned}
\end{equation}

For now we fix $h$ and consider the dual problem of maximization over $\tilde{\mathsf P}$.  For each affine space $\mathsf S^i$ we have defined its dual affine space $\overline{\mathsf S}^i$, whose dual affine space in turn is exactly $\mathsf S^i$ \cite{Chiribella_2016}. Choose an affine basis $\{Q^{i,j}\}_{j=1}^{L_i}$ for $\overline{\mathsf S}^i$, and the maximization problem is further expressed as
\begin{equation}
    \begin{aligned}
    &\max_{\tilde P}\Tr\left[\tilde P\Omega_\phi(h)\right], \\
    &\mathrm{s.t.}\ \,\tilde P=\sum_{i=1}^Kq^iS^i, \sum_{i=1}^Kq^i=1, \\
    &\,q^i\ge0,\ S^i\ge0,\ \Tr\left(S^iQ^{i,j}\right)=1,\ i=1,\dots,K,\ j=1,\dots,L_i.
    \end{aligned}
\end{equation}
Defining $P^i:=q^iS^i$ to avoid the product of variables in optimization, we have
\begin{equation}
    \begin{aligned}
    \max&\Tr\left[\tilde P\Omega_\phi(h)\right], \\
    \mathrm{s.t.}\ &\,\tilde P=\sum_{i=1}^KP^i, \\
    &\sum_{i=1}^Kq^i=1, \\
    &\,P^i\ge0,\ \Tr\left(P^iQ^{i,j}\right)=q^i,\ i=1,\dots,K,\ j=1,\dots,L_i,
    \end{aligned}
\end{equation}
where the constraints $q^i\ge0$ can be safely removed, since $\Tr S^i=\prod_{j=1}^{N}d_{2j}$, implying that $\overline{\mathsf S}^i$ includes a positive operator proportional to identity for any $i=1,\dots,K$, having denoted $d_j := \dim(\map H_j)$ for simplicity. The Lagrangian of the problem is given by 
\begin{equation}
    \begin{aligned}
    L &= \sum_i\Tr\left[P^i\Omega_\phi(h)\right] + \left(1-\sum_iq^i\right)\lambda + \sum_i\Tr\left(P^i\tilde Q^i\right) + \sum_{i,j}\left[q^i-\Tr\left(P^iQ^{i,j}\right)\right]\lambda^{i,j} \\
    &= \lambda + \sum_i\Tr\left\{P^i\left[\Omega_\phi(h)+\tilde Q^i-\sum_j\lambda^{i,j}Q^{i,j}\right]\right\} + \sum_i\left[q^i\left(\sum_j\lambda^{i,j}-\lambda\right)\right],
    \end{aligned}
\end{equation}
for $\tilde Q^i\ge0$. Hence, by removing $\tilde Q^i$ the dual problem is written as
\begin{equation}
    \begin{aligned}
    \min&\,\lambda, \\
    \mathrm{s.t.}\ &\sum_j\lambda^{i,j}Q^{i,j}\ge\Omega_\phi(h),\ \lambda=\sum_{j}\lambda^{i,j},\ i=1,\dots,K,\ j=1,\dots,L_i.
    \end{aligned}
\end{equation}
We define $Q^i := \sum_j\lambda^{i,j}Q^{i,j}/\lambda$ if $\lambda\neq0$ ($\lambda=0$ corresponds to a trivial case where the QFI is zero), and clearly $Q^i$ is an arbitrary operator in the set $\overline{\mathsf S}^i$. Therefore, we cast the dual problem into
\begin{equation}
    \begin{aligned}
    \min&\,\lambda, \\
    \mathrm{s.t.}\ &\,\lambda Q^i \ge \Omega_\phi(h),\ Q^i\in \overline{\mathsf S}^i,\ i=1,\dots,K.
    \end{aligned}
\end{equation}
To further formulate the problem as an SDP, we introduce a block matrix
\begin{equation} 
    A^i = \left( 
\begin{array}{c | c} 
  \frac \lambda 4 I\left(r\times r\right) & 
  \begin{array}{c}
       \<\overline{\dot{\tilde{C}}_{\phi,1}}| \\
       \vdots \\
       \<\overline{\dot{\tilde{C}}_{\phi,r}}|
  \end{array}\\ 
  \hline 
  \begin{array}{ccc}
       |\overline{\dot{\tilde{C}}_{\phi,1}}\> & \hdots & |\overline{\dot{\tilde{C}}_{\phi,r}}\>
  \end{array} & Q^i
 \end{array} 
\right),
\end{equation}
where $|\dot{\tilde{C}}_{\phi,i}\>=|\dot{C}_{\phi,i}\>-\mathrm i\sum_j|C_{\phi,j} \> h_{ji}$. By Schur complement lemma \cite[Theorem 1.12]{Horn2005}, the constraint $\lambda Q^i \ge \Omega_\phi(h)$ is equivalent to the positive semidefiniteness of $A^i$. Then the dual problem is rewritten as
\begin{equation}
    \begin{aligned}
    \min&\,\lambda, \\
    \mathrm{s.t.}\ &\,A^i \ge 0,\ Q^i\in \overline{\mathsf S}^i,\ i=1,\dots,K.
    \end{aligned}
\end{equation}
Slater's theorem \cite{watrous2018theory} implies that the strong duality holds, since the QFI is finite and the inequality constraints can be strictly satisfied for a positive semidefinite operator $\Omega_\phi(h)$, by choosing $\lambda Q^i=\mu\lVert\Omega_\phi(h)\rVert I_{1,2,\dots,2N}$ for $\mu>1$ and any $i=1,\dots,K$, having denoted the operator norm by $\lVert \cdot \rVert$.

Finally, by optimizing the choice of $h$ we derive the result of Theorem \ref{thm_qfi_sdp}. \qed

\subsection{An algorithm for an optimal strategy} \label{sec:optimal strategy}
The proof techniques used in this subsection have been similarly employed for identifying an optimal quantum error correction code \cite{PhysRevResearch.2.013235} or an optimal probe state \cite{Zhou2021PRXQ} for quantum metrology. By solving the SDP for the task QFI in Theorem \ref{thm_qfi_sdp}, we can meanwhile obtain the optimal value of $h$, denoted by $h^{(\mathrm{opt})}$, in 
\begin{equation}
    \min_{h:h^\dag=h}\max_{\tilde{P}\in\widetilde{\strat}}\Tr\left[\tilde{P}\Omega_\phi(h)\right].
\end{equation}
We will show that, $h^{(\mathrm{opt})}$ is also an optimal solution of $h$ in the original optimization problem
\begin{equation}
    \max_{\tilde{P}\in\widetilde{\strat}}\min_{h:h^\dag=h}\Tr\left[\tilde{P}\Omega_\phi(h)\right],
\end{equation}
from which we can solve for an optimal $\tilde P^{(\mathrm{opt})}$, by making sure $(h^{(\mathrm{opt})},\tilde P^{(\mathrm{opt})})$ is a saddle point for $\Tr\left[\tilde{P}\Omega_\phi(h)\right]$.

To this end, we recall the minimax theorem:
\begin{equation} \label{eq:minimax}
   \min_x \max_y f(x,y) = \max_y \min_x f(x,y) 
\end{equation}
for a function $f(x,y)$ convex in $x$ and concave in $y$. Assume $(x_0, y_1)$ is a solution for the L.H.S. of Eq.~(\ref{eq:minimax}) and $(x_1,y_0)$ is a solution for the R.H.S. of Eq.~(\ref{eq:minimax}). It is easy to see that
\begin{equation}
    f(x_0,y_1) \ge f(x_0,y_0) \ge f(x_1,y_0).
\end{equation}
In view of Eq.~(\ref{eq:minimax}) both equalities hold. Therefore, $(x_0,y_0)$ is an optimal solution of the optimization problem and also a saddle point for $f(x,y)$, i.e., $x_0=\argmin_x{f(x,y_0)}$ and $y_0 = \argmax_y{f(x_0,y)}$. By substituting $x=h$ and $y=\tilde P$, we show that, from solving the SDP for QFI, $h^{(\mathrm{opt})}$ is an optimal solution, and we simply need to find $\tilde P^{(\mathrm{opt})}$ such that $(h^{(\mathrm{opt})},\tilde P^{(\mathrm{opt})})$ is a saddle point. 

Clearly, $\tilde P^{(\mathrm{opt})}$ can be identified by the following maximization problem
\begin{equation}
    \begin{aligned}
        &\max_{\tilde{P}\in\widetilde{\strat}}\Tr\left[\tilde{P}\Omega_\phi(h^{(\mathrm{opt})})\right],\\
        &\mathrm{s.t.}\ \,h^{(\mathrm{opt})}=\argmin_h\Tr\left[\tilde P^{(\mathrm{opt})}\Omega_\phi(h)\right].
    \end{aligned}
\end{equation}
Furthermore, $h^{(\mathrm{opt})}=\argmin_h\Tr\left[\tilde P^{(\mathrm{opt})}\Omega_\phi(h)\right]$ can be satisfied by requiring $\partial_{h_{ij}} \left. \Tr\left[\tilde P^{(\mathrm{opt})}\Omega_\phi(h)\right]\right\rvert_{h=h^{(\mathrm{opt})}} = 0,\ \forall i,j$. The derivatives with respect to complex numbers here should be understood as Wirtinger derivatives \cite{Poincare1899,Wirtinger1927} , where $h_{ij}$ and $h_{ji}=\overline{h_{ij}}$ are regarded as independent variables, and then the rules of complex differentiation are fully analogous to normal differentiation with respect to real variables. This is just a trick for simplifying the calculation, and one can equivalently derive the same result by taking derivatives with respect to independent real variables in the Hermitian matrix $h$. The main idea of Wirtinger calculus is that, for a real-differentiable complex-valued function $f(z,\overline z)$ of a complex number $z=x+\mathrm i y$, one can define
\begin{equation}
    \frac{\partial f(z,\overline z)}{\partial z} = \frac 12\left[\frac{\partial f(x,y)}{\partial x} - \mathrm i \frac{\partial f(x,y)}{\partial y}\right]
\end{equation}
and
\begin{equation}
    \frac{\partial f(z,\overline z)}{\partial \overline{z}} = \frac 12\left[\frac{\partial f(x,y)}{\partial x} + \mathrm i \frac{\partial f(x,y)}{\partial y}\right].
\end{equation}
One can easily verify that the requirement of $\partial_xf(x,y)=\partial_yf(x,y)=0$ is equivalent to $\partial_zf(z,\overline z)=\partial_{\overline z}f(z,\overline z)=0$. The definitions can be generalized to functions of mutiple complex variables. For tutorials on Wirtinger calculus, see, for example, Refs.~\cite{kreutzdelgado2009complex,koor2023short}. Leveraging this tool and combining with Eq.~(\ref{eq:performance_op}), the constraint $\partial_{h_{ij}} \left. \Tr\left[\tilde P^{(\mathrm{opt})}\Omega_\phi(h)\right]\right\rvert_{h=h^{(\mathrm{opt})}} = 0,\ \forall i,j$ can be expressed as
\begin{equation} 
    \Tr \left\{\tilde P^{(\mathrm{opt})}\left[-\mathrm i |C_{\phi,i}\> \left(|\dot{C}_{\phi,j}\>-\mathrm i \sum_k\mathrm |C_{\phi,k}\>h^{(\mathrm{opt})}_{kj}\right)^{\dagger} + \left(|\dot{C}_{\phi,i}\>-\mathrm i \sum_l\mathrm |C_{\phi,l}\>h^{(\mathrm{opt})}_{li}\right)\left(-\mathrm i|C_{\phi,j}\>\right)^\dagger\right]^T\right\} 
        =0,\ \forall i,j.
\end{equation}
 This constraint results in Eq.~(\ref{opt_strategy_condition}) in Algorithm \ref{alg_optimal_strategy}. 

Finally, by definition a purification of $\tilde P^{(\mathrm{opt})}$ is an optimal strategy. We can choose any strategy $P^{(\mathrm{opt})}$ such that $\Tr_F P^{(\mathrm{opt})} = \tilde P^{(\mathrm{opt})}$. 

\subsection{Characterization of dual affine spaces} \label{sec:dual affine spaces}
To apply Theorem \ref{thm_qfi_sdp} for the task QFI, the key is to characterize the dual affine spaces $\{\overline{\mathsf S}^i\}$, given a strategy set $\strat$ satisfying (\ref{req_strategy_set}). In general, finding the dual affine space of an arbitrary affine space can be cumbersome, but fortunately the quantum strategy sets introduced in Sections \ref{subsec_defco_strategy}--\ref{subsec_ico} admit simple forms of dual affine spaces $\{\overline{\mathsf S}^i\}$. Such characterization has partly been given in Refs.~\cite{Chiribella_2016,Bavaresco2021PRL}.  

We start with the parallel strategy set $\mathsf{Par}$, given by Eq.~(\ref{eq:parallel strategy set}). By tracing out the global future space $\mathcal H_F$, $\widetilde{\mathsf{Par}} = \{S \ge 0 \mid S \in \mathsf S_{(\mathsf{Par})}\}$ is the set of positive semidefinite operators $I_{2,4,\dots,2N}\otimes \rho_{1,3,\dots,2N-1}$ for $\Tr \rho_{1,3,\dots,2N-1} = 1$. The dual affine space of $\mathsf S_{(\mathsf{Par})}$ is the set of multipartite quantum channels\footnote{Strictly speaking, the dual affine space here is a subset of quantum channels relaxing the requirement of positive semidefiniteness.}
\begin{equation}
    \overline{\mathsf S}_{(\mathsf{Par})}=\{Q\mid Q^\dagger=Q,\ \Tr Q = \prod_{i=1}^N d_{2i-1},\ _{2,4,\dots,2N}Q = _{1,2,\dots,2N}Q\}.
\end{equation}

For the sequential strategy set $\mathsf{Seq}$ defined by Eq.~(\ref{eq:sequential strategy set}), denoting the marginal set after tracing out the global future by $\widetilde{\mathsf{Seq}}=\{S \ge 0 \mid S \in \mathsf S_{(\mathsf{Seq})} \}$, the dual affine space of $\mathsf S_{(\mathsf{Seq})}$ is the set of quantum combs with reversed input-output pairs \cite{Chiribella2009PRA}
\begin{equation}
    \overline{\mathsf S}_{(\mathsf{Seq})}=\{Q\mid Q^\dagger=Q,\ \Tr Q = \prod_{i=1}^N d_{2i-1},\ _{2N,\dots,2i,2i-1} Q= _{2N,\dots,2i} Q\ \text{for}\ i=1,\dots,N\}.
\end{equation}

By definition a general indefinite-causal-order strategy transforms $N$ quantum channels into a quantum channel. By linearity, it can also be legitimately concatenated with any convex combination of $N$ local quantum channels. In fact, a sufficient and necessary condition for characterizing general indefinite-causal-order processes is the requirement of transforming any multipartite no-signaling channels into a global future state \cite{Chiribella2013PRA}. The dual affine space of $\mathsf S_{(\mathsf{ICO})}$ appearing in $\widetilde{\mathsf{ICO}}=\{S \ge 0 \mid S \in \mathsf S_{(\mathsf{ICO})} \}$ is thus the set of no-signaling channels \cite[Definition 6]{Chiribella_2016}
\begin{equation}
    \overline{\mathsf S}_{(\mathsf{ICO})}=\{Q\mid Q^\dagger=Q,\ \Tr Q = \prod_{i=1}^N d_{2i-1},\ _{2i,2i-1} Q= _{2i} Q\ \text{for}\ i=1,\dots,N\}.
\end{equation}

In terms of the strategy sets $\mathsf{SWI}$ and $\mathsf{Sup}$ concerning the superposition of causal orders, either $\widetilde{\mathsf{SWI}}$ or $\widetilde{\mathsf{Sup}}$ is the convex hull of positive semidefinite operators in $N!$ affine spaces $\mathsf S_{(\mathsf{SWI})}^\pi$ or $\mathsf S_{(\mathsf{Sup})}^\pi$ corresponding to different causal orders. $\mathsf S_{(\mathsf{SWI})}^\pi$ is the set of operators
\begin{equation}
    \rho_{2\pi(1)-1}^{\pi}\left(\otimes_{i=1}^{N-1}|I \kk_{2\pi(i),2\pi(i+1)-1}\bb I|_{2\pi(i),2\pi(i+1)-1}\right) \otimes I_{2\pi(N)}
\end{equation}
satisfying $\Tr \rho_{2\pi(1)-1}^{\pi}=1$, for any permutation $\pi\in S_N$. For each $\pi$ the dual affine space is therefore
\begin{equation}
    \overline{\mathsf S}^{\pi}_{(\mathsf{SWI})}=\{Q\mid Q^\dagger=Q,\ \left(\otimes_{i=1}^{N-1}\bb I |_{2\pi(i),2\pi(i+1)-1}\right)\Tr_{2\pi(N)}Q\left(\otimes_{j=1}^{N-1}|I \kk_{2\pi(j),2\pi(j+1)-1}\right) = I_{2\pi(1)-1}\}.
\end{equation}
Analogously, $\widetilde{\mathsf{Sup}}$ is the convex hull of positive semidefinte operators in $\{\mathsf S_{(\mathsf{Sup})}^\pi\}$, corresponding to each sequential order for permutation $\pi$. The dual affine space of $\mathsf S_{(\mathsf{Sup})}^\pi$ is 
\begin{equation}
    \overline{\mathsf S}^{\pi}_{(\mathsf{Sup})}=\{Q\mid Q^\dagger=Q,\ \Tr Q = \prod_{i=1}^N d_{2i-1},\ _{2\pi(N),\dots,2\pi(i),2\pi(i)-1} Q= _{2\pi(N),\dots,2\pi(i)} Q\ \text{for}\ i=1,\dots,N\}.
\end{equation}

We also illustrate the inclusion relations between the relevant dual affine spaces for different strategy sets in Figure~\ref{fig:relations}\protect\subref{subfig:relations_dual_strategies}, except for $\mathsf S_{(\mathsf{SWI})}^\pi$.

\subsection{Comparison with traditional approaches} \label{sec:comparison with traditional approaches}
In this subsection we focus on quantum channel estimation, a task of great interest in quantum metrology. Given $N$ queries to a quantum channel $\mathcal E_\phi$, in our metrology task [cf.~Def.~\ref{defi-metro-task}] we have $C_\phi=E_\phi^{\otimes N}$, where $E_\phi$ is the Choi operator of the channel. Extensive research has been dedicated to the optimization of strategies for quantum channel estimation. We divide these related works into two categories: 
\begin{itemize}
    \item The first type of approaches \cite{Fujiwara2008,Escher2011,Demkowicz-Dobrzanski2012,Demkowicz-Dobrzanski14PRL,Zhou2021PRXQ,kurdzialek2023using,zhou2024limitsnoisyquantummetrology} derive bounds on the QFI based on single-shot quantities of the channel (e.g., the Kraus operators) and circumvent the complexity of optimization for large $N$. Asymptotically tight bounds on QFI (in the limit of $N\to\infty$) have been established for parallel, sequential and causal superposition strategies. In the asymptotic limit, the optimal performance of these strategies coincides with each other \cite{kurdzialek2023using}, and can be achieved by quantum error correction \cite{Zhou2021PRXQ}. For finite $N$, however, these bounds are no longer tight and quantum error correction may not attain the optimal metrological performance \cite{Liu23PRLoptimal,kurdzialek2023using}.
    \item There has also been another type of approaches focusing on the numerical optimization of metrological strategies. Some of these approaches can be found in a review article \cite{Liu22AQT}, and a computing toolkit \cite{Zhang22PRR} including many optimization methods has been implemented. Nevertheless, traditional approaches usually tackle state optimization or control optimization separately rather than fully optimizing the \emph{strategy} as a whole, and the strict optimality of these approaches is not guaranteed.
\end{itemize}

Compared with these traditional approaches, our method provides a strictly optimal solution to the best achievable estimation precision and the strategy that attains it. A drawback of our approach is the growing computational complexity of SDP as $N$ increases, unlike the computation of asymptotically tight bounds which only depend on single-shot channel quantities. Nonetheless, in Ref.~\cite{Liu23PRLoptimal} we showed that, when the set of strategies admits the permutation symmetry, we can significantly reduce the complexity of SDP (at least by an exponential factor) in both QFI evaluation (Theorem \ref{thm_qfi_sdp}) and optimal strategy identification (Algorithm \ref{alg_optimal_strategy}). In Table \ref{tab:SDP complexity} we summarize for each strategy set the SDP complexity in terms of the number of variables concerned, with or without exploiting the symmetry reduction, and compare our approach with traditional methods. Interested readers may refer to Ref.~\cite{Liu23PRLoptimal} for the detailed proof.

\begin{table}[!htbp]
\centering
\begin{tabular}{ccccccc} \toprule
Task & & $\mathsf{Par}$ & $\mathsf{Seq}$ & $\mathsf{SWI}$ & $\mathsf{Sup}$ & $\mathsf{ICO}$ \\
\midrule
\multirow{2}{*}{QFI evaluation}& Ori. & $O\left(s^{2N}\right)$ & $O\left(d^{2N}\right)$ & $O\left(s^{2N}\right)$ & $O\left(N!\,d^{2N}\right)$ & $O\left(d^{2N}\right)$ \\
& Inv. & $O\left(N^{d^2-1}\right)$ & $O\left(d^{2N}\right)$ & $O\left(N^{s^2-1}\right)$ & $O\left(d^{2N}\right)$ & $O\left(N^{d^2-1}\right)$\\
\multirow{2}{*}{Optimal strategy identification} & Ori. & $O\left(\max(s,d_1)^{2N}\right)$ & $O\left(d^{2N}\right)$ & $O\left(N!\right)$ & $O\left(N!\,d^{2N}\right)$ & $O\left(d^{2N}\right)$ \\ 
& Inv. & $O\left(N^{d^2-1}\right)$ & $O\left(d^{2N}\right)$ & $O(N^{s^2-1})$ & $O\left(d^{2N}\right)$ & $O\left(N^{d^2-1}\right)$ \\
QFI upper bound evaluation &  & $O\left(s^2\right)$ \cite{Fujiwara2008,Escher2011,Demkowicz-Dobrzanski2012} & $O\left(s^{2}\right)$ \cite{Demkowicz-Dobrzanski14PRL,kurdzialek2023using} & unknown & $O\left(s^{2}\right)$ \cite{kurdzialek2023using} & unknown\\
\bottomrule
\end{tabular}
\caption{\label{tab:SDP complexity} Computational complexity (in terms of the number of scalar variables) of the SDP in QFI evaluation (Theorem \ref{thm_qfi_sdp}), optimal strategy identification (Algorihtm \ref{alg_optimal_strategy}), and the traditional methods of evaluating QFI upper bounds based on single-shot channel quantities. The QFI upper bounds are asymptotically tight and the computational cost does not scale with $N$, but the bounds can be loose for finite $N$. The numbers of variables in optimization are also compared between the original (Ori.) SDP and the symmetry reduced group-invariant (Inv.) SDP. We have used the notations $d:=d_1 d_2$ for $d_i:=\dim(\mathcal H_i)$ and $s:=\max_\phi\rank(E_\phi)\le d$.}
\end{table}

\section{Applications} \label{sec:applications}

\subsection{Strictly optimal noisy quantum metrology}
Now we present some examples to show the applicability of our theoretical framework for optimal quantum metrology. 
\subsubsection{Amplitude damping and bit flip noise}
As a prototypical example in metrology, we would like to estimate the phase parameter $\phi$ by an optimal strategy in a specified quantum strategy set, given $N$ uses of a noisy quantum channel $\mathcal E_\phi = \mathcal R_z(\phi)\circ \mathcal N$\footnote{Here we take the convention that signal comes after noise as in Ref.~\cite{kurdzialek2023using}, which is different from the assumption in Ref.~\cite{Liu23PRLoptimal}.}, where $\mathcal R_z(\phi)[\cdot]=R_z(\phi)[\cdot] R_z(\phi)^\dagger$ for $R_z(\phi)=e^{-\mathrm i\phi Z/2}$, and $\mathcal N$ is a quantum channel characterizing the noise. We consider two types of noise of great interest in quantum metrology: (1) the amplitude damping channel $\mathcal N^{(\mathrm{AD})}$ described by Kraus operators 
\begin{equation} \label{eq:AD noise}
    K_1^{(\mathrm{AD})}=|0\>\<0|+\sqrt{1-p}|1\>\<1|,\ K_2^{(\mathrm{AD})}=\sqrt{p}|0\>\<1|,
\end{equation}
and (2) the bit flip noise channel $\mathcal N^{(\mathrm{BF})}$ described by Kraus operators 
\begin{equation} \label{eq:BF noise}
    K_1^{(\mathrm{BF})}=\sqrt{1-p}I,\ K_2^{(\mathrm{BF})}=\sqrt{p}X.
\end{equation}

\begin{figure} [!htbp]
\centering
\subfigure[$N=2$, amplitude damping noise.]{\includegraphics[width=0.45\textwidth]{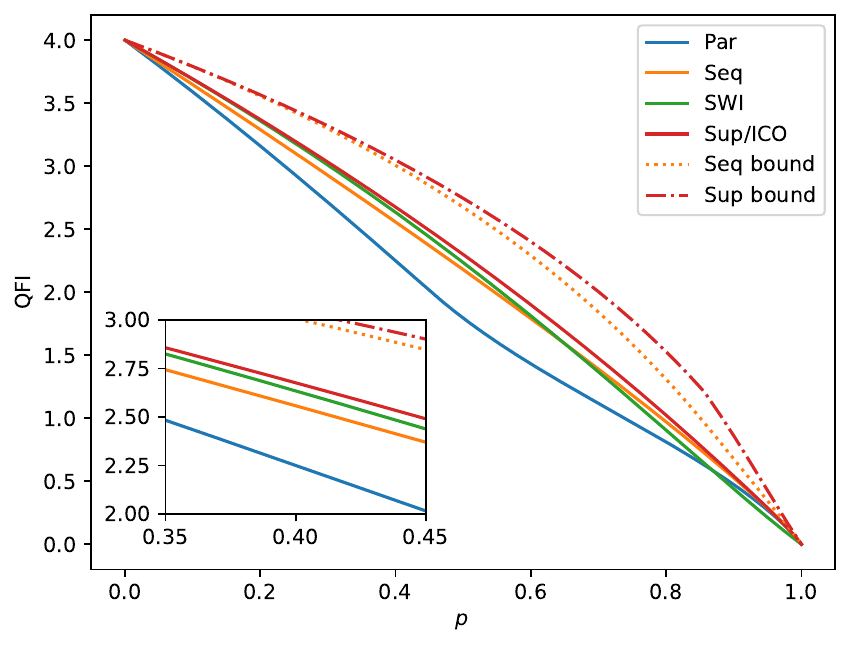}\label{subfig:QFI_AD_hierarchy_N_2}}
\subfigure[$N=3$, amplitude damping noise.]{\includegraphics[width=0.45\textwidth]{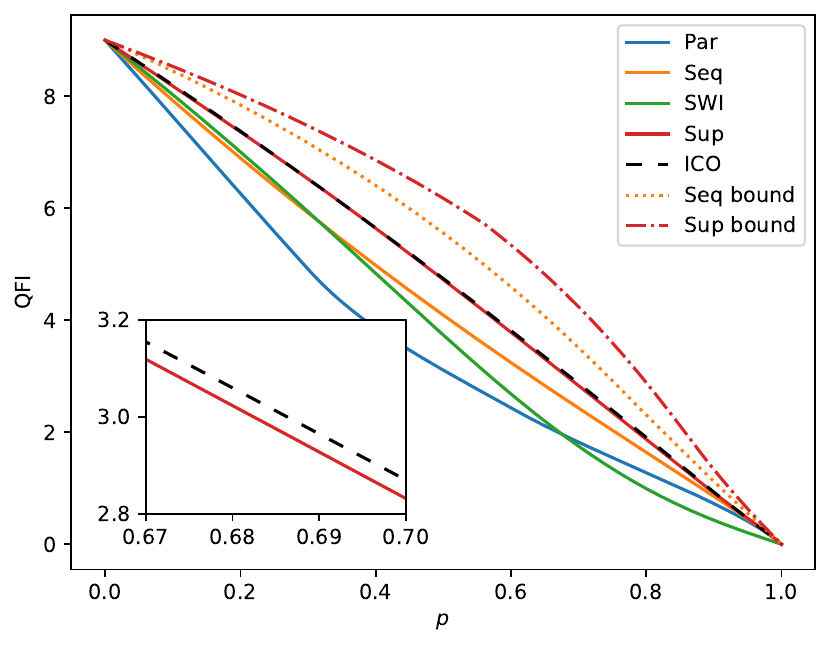}\label{subfig:QFI_AD_hierarchy_N_3}}
\subfigure[$N=2$, bit flip noise.]{\includegraphics[width=0.45\textwidth]{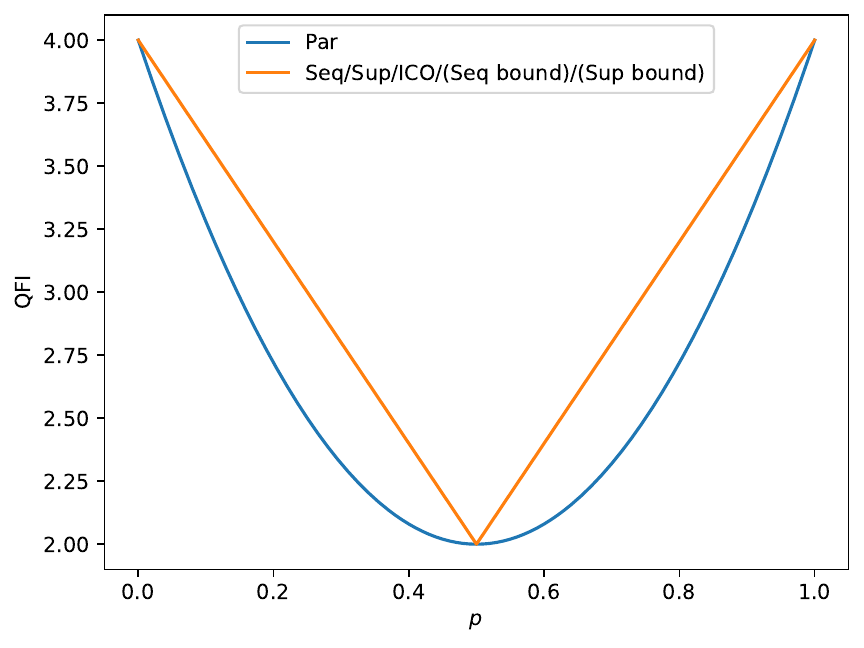}\label{subfig:QFI_X_hierarchy_N_2}}
\subfigure[$N=3$, bit flip noise.]{\includegraphics[width=0.45\textwidth]{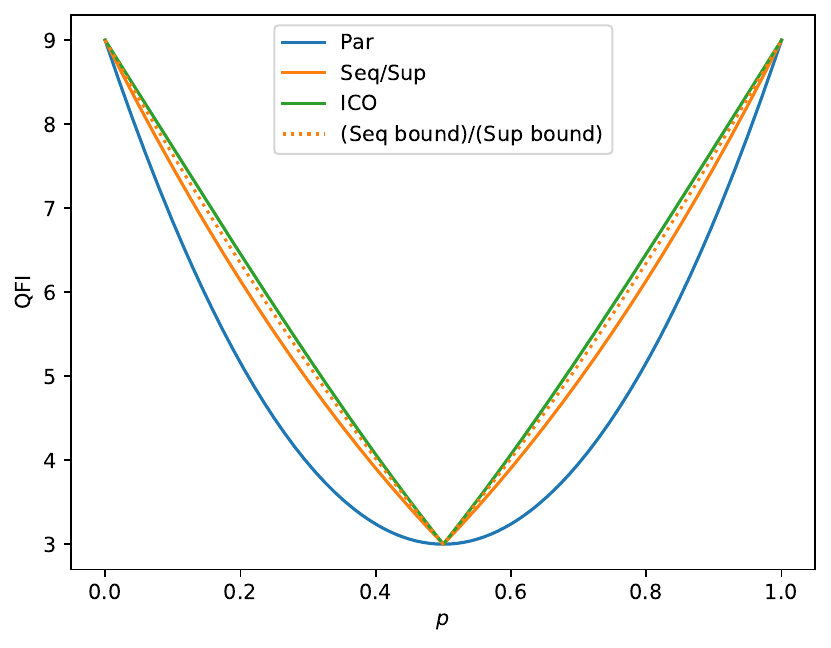}\label{subfig:QFI_X_hierarchy_N_3}}
\caption{\label{fig:QFI hierarchy}Optimal QFI versus the noise strength $p$ with different strategy sets. Inset figures magnify certain parts of the figures to provide a better illustration of the gaps between the QFI of different strategies. The solid or dashed lines represent the exact QFI (up to a negligible numerical error) yielded by our approach, and the dotted or dashdotted lines represent the state-of-the-art upper bounds on QFI obtained by Algorithm 2 in Ref.~\cite{kurdzialek2023using}. Different lines are plotted to coincide with each other with a common label up to a numerical error of $10^{-6}$, for example, ``$\mathsf{Sup}/\mathsf{ICO}$'' in \protect\subref{subfig:QFI_AD_hierarchy_N_2}.}
\end{figure}

The two types of noise considered here result in two different optimal precision scalings in quantum metrology. It has been known that, using asymptotically optimal parallel, sequential or causal superposition strategies, the QFI of phase estimation with amplitude damping noise follows the standard quantum limit (SQL) $J^{(N)}(\mathcal E_\phi)=\Theta(N)$ and phase estimation with bit flip noise follows the Heisenberg limit (HL) $J^{(N)}(\mathcal E_\phi)=\Theta(N^2)$\footnote{Remarkably, however, a super-Heisenberg scaling of quantum metrology with the quantum SWITCH can been demonstrated in continuous-variable (infinite-dimensional) systems with bounded energy \cite{Zhao2020PRL,Yin2023NP}.} \cite{Escher2011,Demkowicz-Dobrzanski2012,Demkowicz-Dobrzanski14PRL,Zhou2021PRXQ,kurdzialek2023using}, where $J^{(N)}(\mathcal E_\phi)$ denotes the QFI one can obtain by concatenating $N$ channels $\mathcal E_\phi$ with an optimal strategy. These three types of strategies always have the same asymptotic performance in quantum channel estimation. Our results indicate that, however, in both cases of SQL and HL, it is possible to obtain metrological advantages by using sequential strategies and indefinite causal order for a finite $N$. 

We plot the optimal QFI of $N=2$ or $3$ uses of the channel versus the noise strength $p$ with different strategy constraints at $\phi=\pi/2$ in Figure~\ref{fig:QFI hierarchy}. For the amplitude damping noise, when $N=2$ we observe $J^{(\mathsf{Par})}<J^{(\mathsf{Seq})}<J^{(\mathsf{Swi})}<J^{(\mathsf{Sup})}=J^{(\mathsf{ICO})}$ (up to a numerical error of no more than $10^{-6}$) at $p=0.4$, and when $N=3$ we find a full strict hierarchy $J^{(\mathsf{Par})}<J^{(\mathsf{Seq})}<J^{(\mathsf{Swi})}<J^{(\mathsf{Sup})}<J^{(\mathsf{ICO})}$ at $p=0.2$. Remarkably, a simple quantum SWITCH strategy, which does not require any intermediate control between estimated channels, can sometimes outperform an optimized sequential strategy with arbitrary control\footnote{For example, for $N=2$ and $p=0.4$ with the amplitude damping noise, $J^{(\mathsf{Swi})}/J^{(\mathsf{Seq})}$=1.03.}. This highlights the possibility that an advantage of the quantum SWITCH over \emph{any} sequential strategy can be demonstrated in the lab, bearing in mind the high circuit complexity typically required for implementing the intermediate control (see Section \ref{sec:circuit decomposition theory}). For the bit flip noise, the problem is symmetric ($X\leftrightarrow I$) with respect to the reflection against $p=0.5$, which coincides with the plotted lines in Figures \ref{fig:QFI hierarchy}\protect\subref{subfig:QFI_X_hierarchy_N_2} and \protect\subref{subfig:QFI_X_hierarchy_N_3}. When $0<p<1$ and $p\neq 0.5$, we obtain $J^{(\mathsf{Par})}<J^{(\mathsf{Seq})}=J^{(\mathsf{Sup})}=J^{(\mathsf{ICO})}$ when $N=2$ and $J^{(\mathsf{Par})}<J^{(\mathsf{Seq})}=J^{(\mathsf{Sup})}<J^{(\mathsf{ICO})}$ when $N=3$. The general indefinite causal order $\mathsf{ICO}$ exhibits a larger advantage over $\mathsf{Sup}$ for $N=3$ in this example.

In Figure~\ref{fig:QFI hierarchy}, we also compare the exact values of the optimal QFI (up to a negligible numerical error) obtained by Theorem \ref{thm_qfi_sdp} and the state-of-the-art upper bounds (expressed in Kraus operators of a single channel and their derivatives) on the QFI with sequential and causal superposition strategies computed by Algorithm 2 in Ref.~\cite{kurdzialek2023using}. An interesting observation (which has also been noted by Ref.~\cite{kurdzialek2023using}) is that, as shown in Figure~\ref{fig:QFI hierarchy}\protect\subref{subfig:QFI_X_hierarchy_N_3}, for $N=3$ channels with the bit flip noise, the exact QFI $J^{(\mathsf{ICO})}$ can be strictly larger than the upper bound on $J^{(\mathsf{Sup})}$, which certainly implies that the upper bound on $J^{\mathsf{Sup}}$ does not apply to the most general indefinite-causal-order strategies. This leaves room for exploring whether or not $\mathsf{ICO}$ can provide any asymptotic advantage in channel estimation. On the other hand, little is known on whether and how strategies in $\mathsf{ICO}$ can be physically realized.

\subsubsection{Noisy metrology in NMR experiments}
To connect our theoretical framework more closely to the real-world physical realization, we further investigate a noise model often encountered in nuclear magnetic resonance (NMR) experiments. For single qubits, the environment-induced decoherence is typically characterized by the longitudinal relaxation time $T_1$ and the transverse relaxation time $T_2$. The decoherence of a qubit can be phenomenologically characterized by the density matrix transformation \cite{Nielsen_Chuang_2010}
\begin{equation}
    \left(\begin{array}{cc}
         a & b \\
         \overline{b} & 1-a
    \end{array}\right) \rightarrow \left(\begin{array}{cc}
         (a-a_0)e^{-t/T_1}+a_0 & be^{-t/T_2} \\
         \overline{b}e^{-t/T_2} & (a_0-a)e^{-t/T_1}+1-a_0
    \end{array}\right),
\end{equation}
where the parameter $a_0$ characterizes the equilibrium state and $t$ is the evolution time. Such a process can be modelled by Kraus operators
\begin{equation} \label{eq:NMR noise}
    \begin{aligned}
        K_1^{(\mathrm{NMR})}&=\sqrt{1-\alpha}|1\>\<0|,\ K_2^{(\mathrm{NMR})}=\sqrt{1-\beta}|0\>\<1|, \\
        K_3^{(\mathrm{NMR})}&=\frac{1}{\sqrt{2}}\left\{\frac{\alpha+\beta-\sqrt{\gamma^2+(\alpha-\beta)^2}}{\gamma^2+\left[\alpha-\beta-\sqrt{\gamma^2+(\alpha-\beta)^2}\right]^2}\right\}^{1/2}\left\{\left[\alpha-\beta-\sqrt{\gamma^2+(\alpha-\beta)^2}\right]|0\>\<0|+\gamma|1\>\<1|\right\},\\
        K_4^{(\mathrm{NMR})}&=\frac{1}{\sqrt{2}}\left\{\frac{\alpha+\beta+\sqrt{\gamma^2+(\alpha-\beta)^2}}{\gamma^2+\left[\alpha-\beta+\sqrt{\gamma^2+(\alpha-\beta)^2}\right]^2}\right\}^{1/2}\left\{\left[\alpha-\beta+\sqrt{\gamma^2+(\alpha-\beta)^2}\right]|0\>\<0|+\gamma|1\>\<1|\right\},
    \end{aligned}
\end{equation}
where $\alpha=(1-a_0)e^{-t/T_1}+a_0$, $\beta=a_0e^{-t/T_1}+1-a_0$, and $\gamma=2e^{-t/T_2}$.

\begin{figure} [!htbp]
\centering
\subfigure[$N=2$.]{\includegraphics[width=0.45\textwidth]{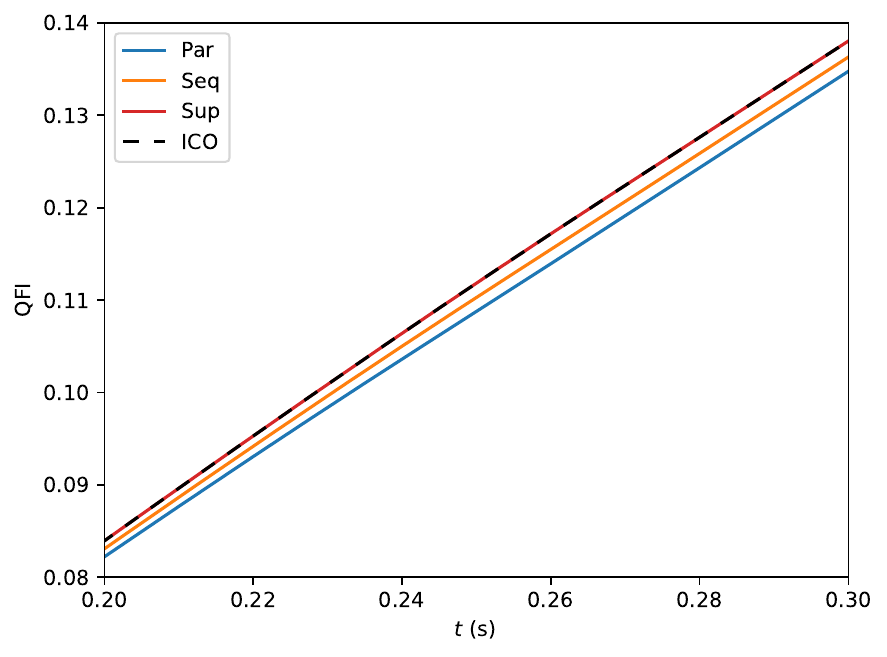}\label{subfig:QFI_NMR_hierarchy_N_2}}
\subfigure[$N=3$.]{\includegraphics[width=0.45\textwidth]{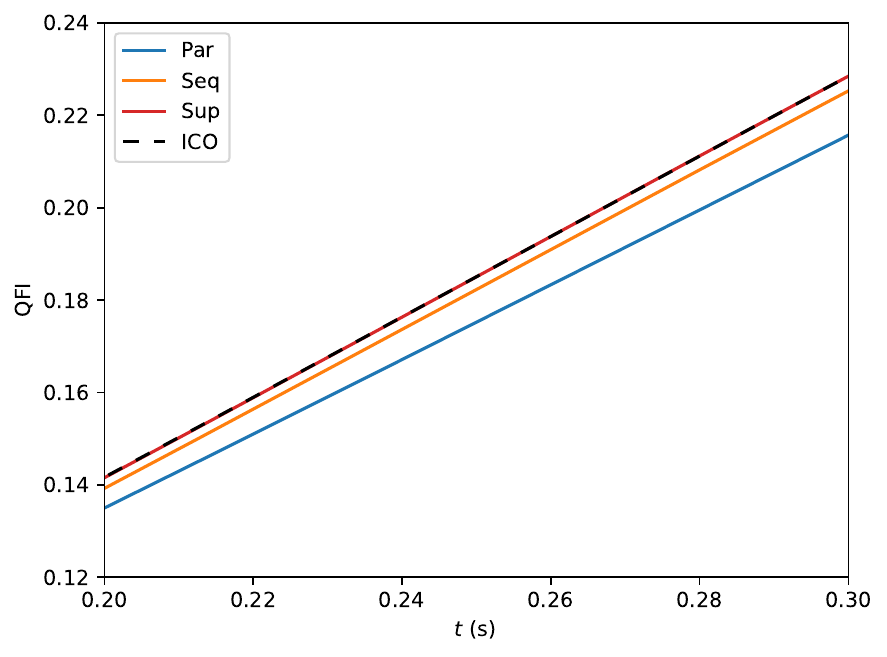}\label{subfig:QFI_NMR_hierarchy_N_3}}
\caption{\label{fig:QFI hierarchy NMR}Optimal QFI versus the evolution time $t$ with different strategy sets under the NMR experimental noise.}
\end{figure}

Now we study the optimal performance of different metrological strategies for frequency estimation under such decoherence noise model. Concretely, we would like to estimate $\omega$ from $N$ uses of $\mathcal E_\omega=\mathcal U_z(\omega)\circ \mathcal N^{(\mathrm{NMR})}$, where $\mathcal U_z(\omega)[\cdot] = e^{-\mathrm i\omega t Z/2}[\cdot]e^{\mathrm i\omega t Z/2}$ and $\mathcal N^{(\mathrm{NMR})}$ is characterized by the Kraus operators in Eq.~(\ref{eq:NMR noise}). In accordance with typical NMR experimental noise characteristics \cite{Long22PRLentanglement}, we take $T_1=3.2\mathrm{s}$, $T_2=1.1\mathrm{s}$ and $a_0=0.5$, implying that the qubit state equilibrates to the maximally mixed state in the long-time limit $t\gg T_{1}$ and $t\gg T_{2}$. In Fig.~\ref{fig:QFI hierarchy NMR}, we plot the evolution of QFI versus $t$ for $N=2$ or $3$ with different strategies, taking the ground truth $\omega=10\mathrm{kHz}$. Under this noise model, we also observe a hierarchy $J^{(\mathsf{Par})}<J^{(\mathsf{Seq})}<J^{(\mathsf{ICO})}$ with small gaps, which implies that parallel strategies are nearly optimal even compared to the most general indefinite-causal-order strategies. The optimal performance of causal superposition strategies almost coincides with that of general indefinite-causal-order strategies (the latter has a negligible advantage when $N=3$). 

\subsubsection{Estimation of non-identical channels}
\begin{figure} [!htbp]
    \centering
    {\includegraphics[width=0.45\textwidth]{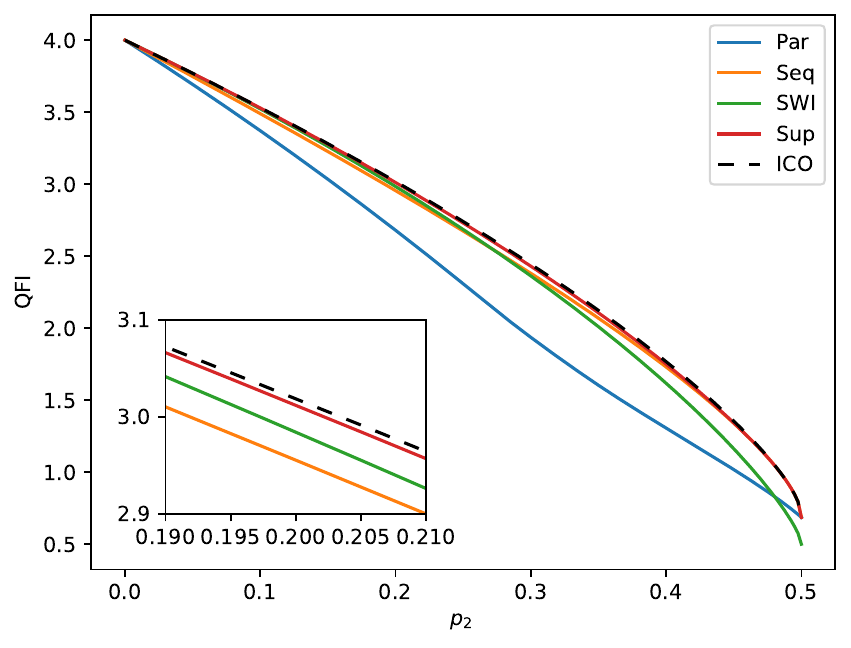}}
    \caption{Comparison of the QFI with different strategy constraints versus the amplitude damping noise strength $p_2 \in [0,0.5]$, assuming $p_1=2p_2$. We zoom in on the interval $p\in[0.19,0.21]$          to exhibit strict gaps between all the strategies.}
    \label{fig:QFI_different_AD_strategies}
\end{figure}

Having established a strict hierarchy for quantum metrology for $N$ uses of the same quantum channel, we turn to the problem of estimating non-identical channels under our theoretical framework. Assume we are given two channels $\mathcal E_\phi^{(1)} = \mathcal R_z(\phi)\circ \mathcal N^{(1)}$ and $\mathcal E_\phi^{(2)} = \mathcal R_z(\phi)\circ \mathcal N^{(2)}$, where $\mathcal N^{(1)}$ and $\mathcal N^{(2)}$ are noisy channels with different noise strength $p_1$ and $p_2$. We take the same amplitude damping noise model as defined by Eq.~(\ref{eq:AD noise}) for example, but with the assumption $p_1=2p_2$, making the problem non-symmetric. The QFI with different strategy constraints for $N=2$ and $\phi=\pi/2$ is plotted in Figure~\ref{fig:QFI_different_AD_strategies}, where $J^{(\mathsf{Seq})}$ is defined as the maximal QFI value of the two sequential orders. Similar to the case of estimating identical channels with amplitude damping noise, we identify a strict hierarchy between the QFI with all the five sets of strategies.

\subsubsection{Benchmarking the performance of existing protocols}
\begin{figure} [!htbp]
    \centering
    {\includegraphics[width=0.45\textwidth]{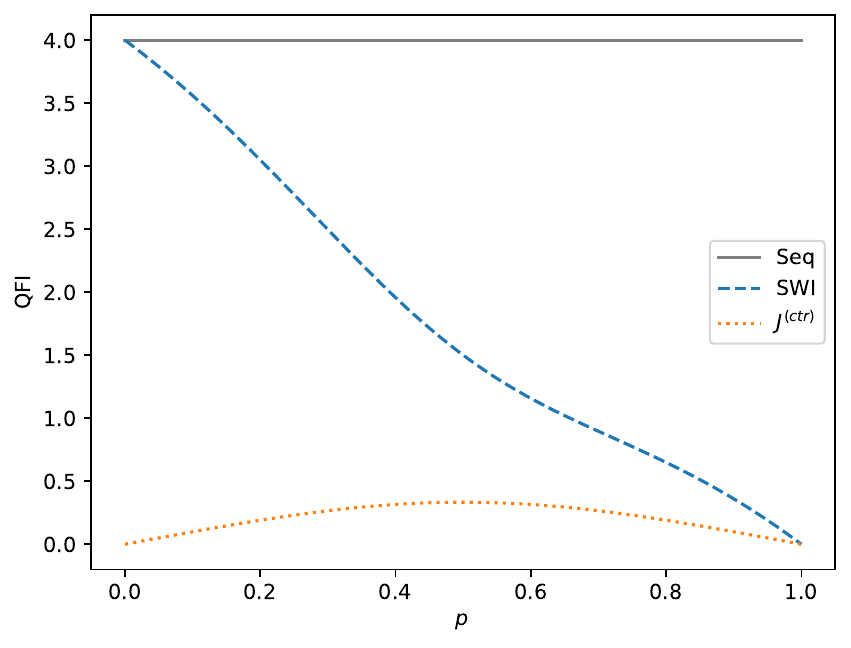}}
    \caption{Comparison of the QFI $J^{(\mathrm{ctr})}$ obtained by the protocol in Ref.~\cite{An24PRAnoisy} and the optimal QFI $J^{(\mathsf{SWI})}$ as well as $J^{(\mathsf{Seq})}$, versus the phase flip noise strength $p$.}
    \label{fig:QFI_signalX_PhaseFlip_strategies}
\end{figure}

Our theoretical tool can be applied to provide rigorous benchmarking for experiments in quantum metrology. For instance, Ref.~\cite{An24PRAnoisy} experimentally demonstrated the performance of the quantum SWITCH in phase estimation with noisy general Pauli channels, following the theoretical investigation in Ref.~\cite{Chapeau-Blondeau2021PRA}. Refs~\cite{Chapeau-Blondeau2021PRA,An24PRAnoisy} evaluated the Fisher information when only the control qubit is measured, and using our approach we can readily benchmark this protocol by computing the QFI of an optimal quantum SWITCH strategy, which in general can be achieved by a joint measurement on the output state. It is worth noting that sometimes only measuring the control qubit of the quantum SWTICH can be far from optimal. Apparently, in the noiseless case where identical unitary channels commute, measuring the control qubit yields no information about the parameter, while measuring the system state attains the maximal QFI. Following the setup in Ref.~\cite{An24PRAnoisy}, consider the problem of estimating two copies of $\mathcal E_\phi = \mathcal N^{(\mathrm{PF})} \circ \mathcal R_x(\phi)$, where $\mathcal R_x(\phi)[\cdot] = e^{-\mathrm i\phi X/2}[\cdot]e^{\mathrm i\phi X/2}$ encodes the parameter of interest $\phi$, and $\mathcal N^{(\mathrm{PF})}(\rho) = (1-p)\rho + p Z\rho Z$ is the phase flip noise channel. As illustrated in Figure~\ref{fig:QFI_signalX_PhaseFlip_strategies}, for $\phi=\pi/2$, measuring the control qubit yields the QFI $J^{(\mathrm{ctr})} = 1/3$ when $p=0.5$ \cite[Eq.~(31)]{An24PRAnoisy} (the highest QFI for all $p$), which is $77.8\%$ lower than the highest QFI $J^{(\mathsf{SWI})}=1.5$ obtained by an optimal quantum SWTICH strategy. Meanwhile, $J^{(\mathrm{ctr})}$ is $91.7\%$ lower than the optimal sequential QFI $J^{(\mathsf{Seq})}=4$---the highest QFI one can obtain without the assistance of indefinite causal order (for example, by applying quantum error correction \cite{Duer2014PRL,Kessler2014PRL,Demkowicz-Dobrza2017PRX,Zhou2018,Zhou2021PRXQ}). Our theoretical framework can be therefore useful for assessing and establishing the advantage of indefinite causal order more rigorously (see Ref.~\cite{mothe2023PRAreassessing} for the analysis of some other examples).

\subsection{Memory effect in non-Markovian quantum metrology}

This formalism can also apply to estimating the non-Markovian multi-step quantum processes when a predetermined causal order is maintained \cite{Altherr2021PRL}. Operationally, a non-Markovian process comprises a sequence of quantum channels with memory (environment), giving rise to temporal correlations \cite{Pollock18PRA}. With the environment typically inaccessible, a non-Markovian process is a sequential process with the accessible inputs and outputs for the system at multiple steps, and mathematically characterized by a quantum comb (see Figure~\ref{fig:gen_comb}). A parametrized non-Markovian process $\mathcal C_\phi$ naturally fits in the metrology task (see Definition \ref{defi-metro-task}) and can thus be tackled in a similar fashion to channel estimation. Nevertheless, it is worth mentioning that such a process is causally definite and should be estimated with strategies of definite causal orders, e.g., parallel and sequential ones.

\begin{figure} [!htbp]
\centering
\subfigure[Non-Markovian process.]{\includegraphics[width=0.45\textwidth]{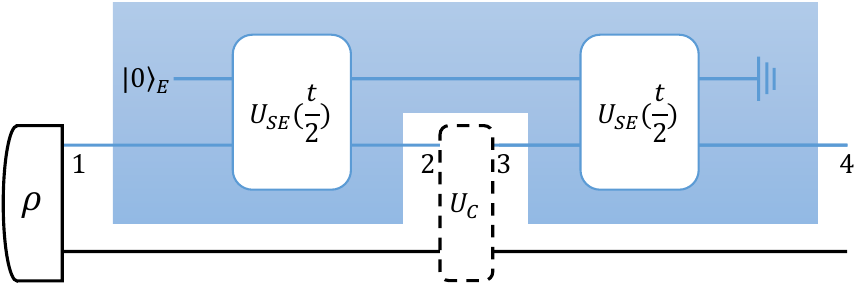}\label{subfig:nonMarkovian_process}}\qquad
\subfigure[Markovian counterpart.]{\includegraphics[width=0.45\textwidth]{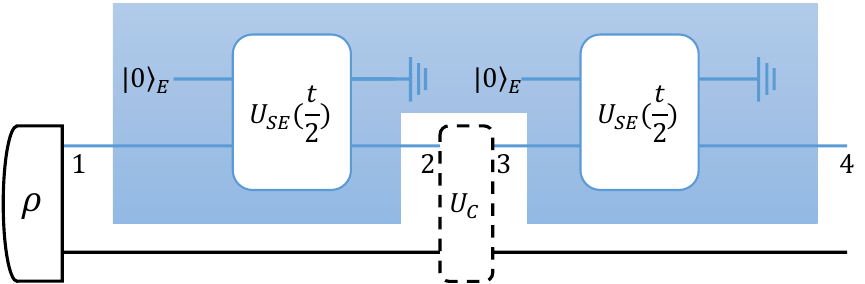}\label{subfig:Markovian_process}}
\caption{\label{fig:nonMarkov_Markov_process}Illustration of a two-step non-Markovian process and its Markovian counterpart. We investigate the estimation of both processes with optimal parallel, sequential and control-free strategies. The non-Markovian process to estimate is encompassed in the blue shaded area, and the strategy is allowed to use ancillae. For parallel, sequential and control-free strategies, $U_c$ can be an ancilla-assisted SWAP gate, arbitrary unitary, and identity channel, respectively.}
\end{figure}

Consider the evolution $U_{SE}(t)=e^{-\mathrm i H_{SE}t}$ with the system-environment Hamiltonian $H_{SE}=H_0+H_1$, where $H_0=\phi Z\otimes I$ encodes the signal $\phi$ on the system qubit, and $H_1=g(X\otimes X+Y\otimes Y+Z\otimes Z)$. $H_1$ generates a SWAP-type interaction between the system and the environment, as $e^{-\mathrm i H_1 t}$ is a SWAP gate when $t=\frac{\pi}{4g}$. We assume that the inaccessible environment is initialized in $|0\>$, and we can apply an intermediate control operation to the system and possible ancillae at time $t/2$, as illustrated in Figure~\ref{fig:nonMarkov_Markov_process}\protect\subref{subfig:nonMarkovian_process}. We allow for fast and accurate control over the system, but cannot have additional control over the environment. To examine the memory effect, we also investigate the Markovian counterpart of the process as depicted in Figure~\ref{fig:nonMarkov_Markov_process}\protect\subref{subfig:Markovian_process}, where the environment is reinitialized in the middle and the information backflow to the system is prohibited. In both scenarios, we compare the optimal QFI obtained by parallel, sequential, and control-free strategies. Here a parallel strategy corresponds to a ``feedforward'' strategy where the output state is never fed back into the unknown process, while a control-free strategy is a special sequential strategy where the control operation is a trivial identity channel.

\begin{figure} [!htbp]
    \centering
    {\includegraphics[width=0.45\textwidth]{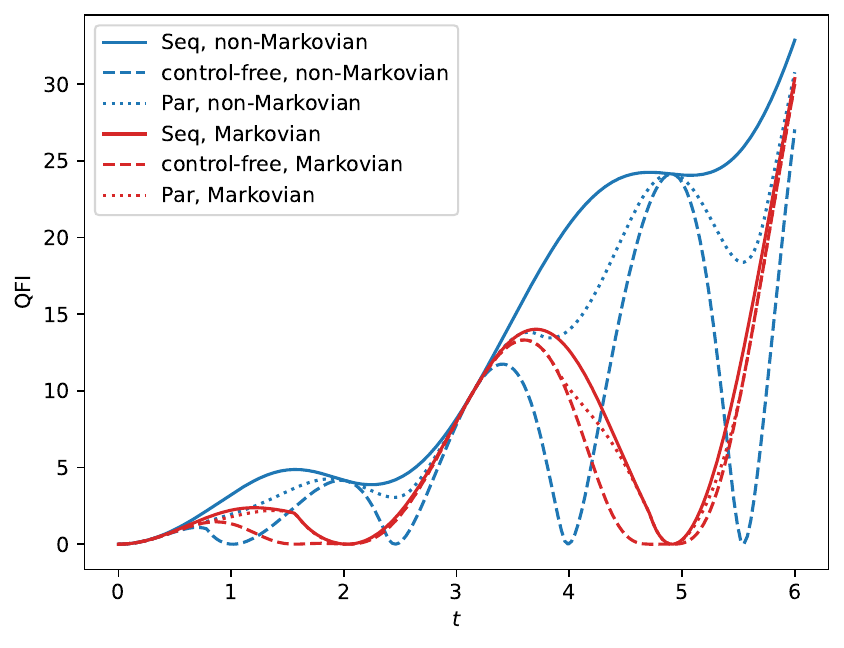}}
    \caption{Comparison of the QFI with parallel, sequential and control-free strategies for non-Markovian metrology. Red and blue lines correspond to non-Markovian and Markovian processes respectively. Furthermore, we use solid, dashed and dotted lines to represent the QFI of sequential, control-free and parallel strategies.}
    \label{fig:QFI_nonMarkov}
\end{figure}

The QFI corresponding to different scenarios for $\phi=0$ and $g=1.0$ is plotted in Figure~\ref{fig:QFI_nonMarkov}. The oscillating behaviour of the QFI apparently arises from the SWAP-type interaction between the system and the environment. Using an optimal sequential strategy---the best strategy following a definite causal order, the non-Markovian process (blue solid line) yields a significantly higher QFI than the Markovian counterpart (red solid line), which signifies the information flow back to the system from the environment. It is easy to seem that the estimation of the Markovian counterpart is equivalent to quantum channel estimation. The intermediate control operation is also important, otherwise one would obtain no information at certain time. The memory effect can be similarly manifested when we use a parallel strategy, as a large gap can be identified between the QFI of the non-Markovian and Markovian process.

We remark that our approach can also be applicable to some cases when the Hilbert space of the environment has large or even infinite dimensions, as the dimension of the quantum comb describing the non-Markovian process only depends on the system. This could provide an ultimate precision limit for quantum metrology of finite-dimensional quantum systems in the continuous variable environment, which is ubiquitous, for example, in optical experiments.

\subsection{Designing optimal protocols}

\subsubsection{Optimal protocols by the quantum comb decomposition} \label{sec:optimal protocols comb decomposition}

In this subsection, we present a concrete example to manifest how to implement the optimal strategy yielded by Algorithm \ref{alg_optimal_strategy} by universal quantum gates, including single-qubit gates and CNOT gates, based on the decomposition methods in Section \ref{sec:circuit decomposition theory}. With the freedom of choosing a parameter-independent unitary on the final output state, we can slightly adjust the strategy output by Algorithm \ref{alg_optimal_strategy} to further reduce the CNOT count without affecting the QFI. In terms of an optimal causal superposition strategy, we simply need to follow the comb decomposition routine for each sequential strategy branch in the superposition. Taking into account the permutation symmetry of the problem \cite{Liu23PRLoptimal}, we can choose an optimal causal superposition strategy such that each sequential branch contains the same state preparation and intermediate control, which could facilitate the experimental demonstration. 

\threesubsection{Optimal sequential strategy} \label{app:implementation of seq}
The Choi operator of a sequential strategy $P \in \map L\left(\map H_F\otimes_{i=0}^{2N} \map H_i\right)$ is an $(N+1)$-step quantum comb, where $\map H_0=\mathbb C$ is trivial and $\map H_{2N+1}=\map H_F$ is the global future space. Based on Section \ref{sec:circuit decomposition theory} we can obtain a sequence of isometries $V^{(k)}\in\map L\left(\map H_{2k-1} \otimes \map H_{A_{k-1}}, \map H_{2k} \otimes \map H_{A_{k}}\right)$ with minimal ancilla space of $\dim (\map H_k)=\rank(P^{(k)})$.

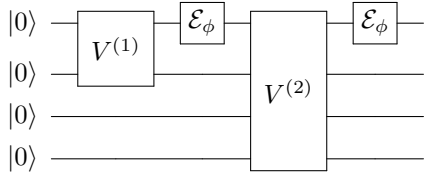
\begin{figure} [!htbp]
    \centering
    \mbox{
    \Qcircuit @C=1em @R=.7em {
\lstick{|0\>} & \multigate{1}{V^{(1)}} & \gate{\map E_\phi} & \multigate{3}{V^{(2)}} & \gate{\map E_\phi} & \qw \\
\lstick{|0\>} & \ghost{V^{(1)}}& \qw & \ghost{V^{(2)}} & \qw & \qw \\
\lstick{|0\>} & \qw & \qw & \ghost{V^{(2)}} & \qw & \qw \\
\lstick{|0\>} & \qw & \qw & \ghost{V^{(2)}} & \qw & \qw 
}
}
    \caption{A sequence of isometries corresponding to a sequential strategy for $N=2$. The first qubit is the system qubit going through the channel $\map E_\phi$ twice, while the three other qubits are ancillary.}
    \label{fig:V1 and V2 seq}
\end{figure}

As the last isometry $V^{(N+1)}$ preserves the QFI, it is only necessary to consider the implementation of the first $N$-step $P^{(N)}$ instead of the full strategy $P^{(N+1)}$. In the case of $N=2$ qubit channels, it is easy to see that $\dim(\map H_{A_{1}})\le 2$ and $\dim(\map H_{A_{2}}) \le 8$, so $V^{(1)}$ is an isometry from $0$ to (at most) $2$ qubits and $V^{(2)}$ is an isometry from $2$ to (at most) $4$ qubits, as illustrated in Figure~\ref{fig:V1 and V2 seq}.  

Next, we apply a circuit decomposition of each isometry into single-qubit gates and CNOT gates. First, $V^{(1)}$ is the preparation of a two-qubit state, which in general requires only one CNOT gate \cite{Znidaric08PRA}. Second, $V^{(2)}$ is an isometry from two to four qubits, and the state-of-the-art decomposition scheme is the column-by-column approach which requires at most 54 CNOT gates \cite{Iten2016PRA}. Furthermore, as an arbitrary unitary on three ancillae can always be absorbed into $V^{(3)}$ and therefore does not affect the QFI, we have the freedom to choose a proper $V^{(2)}$, which can further reduce the worst CNOT count to $47$ without changing the QFI.

As explained in Subsection \ref{sec:circuit decomposition theory}, in the column-by-column decomposition of an isometry $V$ from $m$ to $n$ qubits ($m\le n$) we need to find a sequence of unitary operations $U=U_{2^m-1}\cdots U_0$ that transforms $V$ into $I(2^n\times 2^m)$ column by column, and then apply $U^\dagger$ with $n-m$ ancilla qubits intialized as $|0\>$ for implementing the original isometry. Here we only focus on $U_0$ (the last unitary in implementing $U^\dagger$), which is the inverse of the process preparing a state $V|{0}\>^{\otimes m}$ from $|0\>^{\otimes n}$. In terms of decomposing $V^{(2)}$ from $m=2$ to $n=4$ qubits, preparing a four-qubit state in general requires eight CNOT gates \cite{Plesch11PRA}. Fortunately, without changing the QFI, we have the freedom to choose a unitary $U_{\mathrm{anc}}$ on the ancillae after applying $V^{(2)}$ such that the state $V^{(2)\prime}|0\>^{\otimes 2}=U_{\mathrm{anc}}V^{(2)}|0\>^{\otimes 2}$ can be prepared using only one CNOT gate. This can be seen by dividing the four qubits into two parties, including the single system qubit (in the space $\map H_S$) and the three ancille (in the space $\map H_A$), and taking the Schimidt decomposition of the four-qubit state $V^{(2)}|0\>^{\otimes 2}$
\begin{equation}
    |\psi\>_{SA} := V^{(2)}|0\>^{\otimes 2} = \sum_{i=0}^1 \lambda_i |e_i\>_S |f_i\>_A,
\end{equation}
where $\{|e_i/f_i\>_{S/A}\}$ forms an orthonormal basis of $\map H_{S/A}$, and $\{\lambda_i\}$ is a set of nonnegative real numbers satisfying $\sum_i\lambda_i^2=1$. Therefore, to prepare $V^{(2)}|0\>^{\otimes 2}$, we only need a local unitary on $\map H_S$ to generate $\sum_{i=0}^1 \lambda_i |i\>_S|0\>_A$, then apply one CNOT gate taking the system qubit as the control to obtain $\sum_{i=0}^1 \lambda_i |i\>_S|i\>_A$, and finally apply local unitary operations $U_S=\sum_i|e_i\>_S\<i|_S$ on the system and $U_A=\sum_i|f_i\>_A\<i|_A$ on the ancillae respectively. If we take $V^{(2)\prime}=U_{\mathrm{anc}}V^{(2)}$ where $U_{\mathrm{anc}} = U_A^{\dagger}$, then it is easy to see that $V^{(2)\prime}|0\>^{\otimes 2}=U_{\mathrm{anc}}V^{(2)}|0\>^{\otimes 2}=\sum_{i=0}^1 \lambda_i |e_i\>_S |i\>_A$ can thus be prepared using one CNOT gate. This choice of $V^{(2)\prime}$ saves $7$ CNOT gates compared to the general state preparation scheme, and leads to a worst CNOT count of $47$ in total.

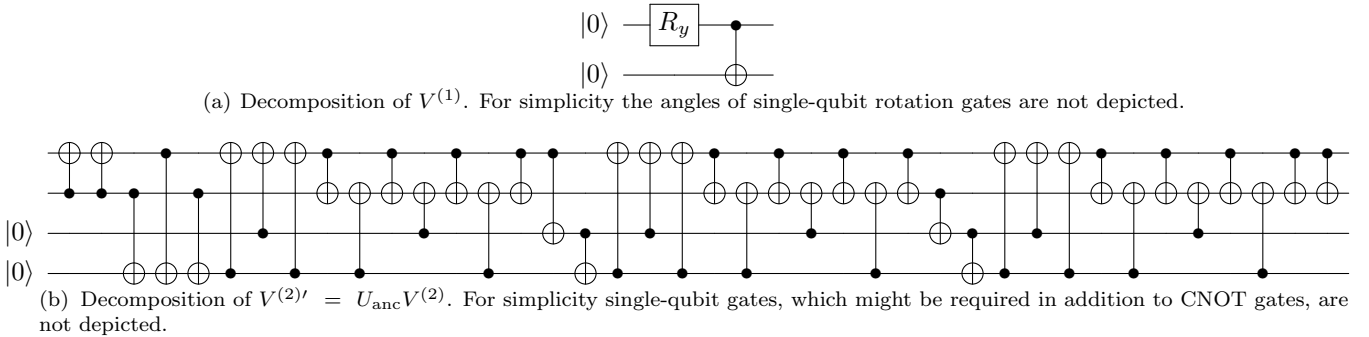
\begin{figure} [!htbp]
\centering
\subfigure[Decomposition of $V^{(1)}$. For simplicity the angles of single-qubit rotation gates are not depicted.]{\mbox{
    \Qcircuit @C=1em @R=.7em {
\push{\rule{20em}{0em}} & \lstick{|0\>}&  \gate{R_y} & \ctrl{1} & \qw & \push{\rule{20em}{0em}} \\
\push{\rule{20em}{0em}} & \lstick{|0\>} & \qw  & \targ & \qw & \push{\rule{20em}{0em}} 
}
}\label{subfig:seq V1 decomp}}\\
\subfigure[Decomposition of $V^{(2)\prime}=U_{\mathrm{anc}}V^{(2)}$. For simplicity single-qubit gates, which might be required in addition to CNOT gates, are not depicted.]{\mbox{
    \Qcircuit @C=.4em @R=.7em {
 & \targ & \targ & \qw & \ctrl{3} & \qw & \targ & \targ & \targ & \ctrl{1} & \qw  & \ctrl{1} & \qw  & \ctrl{1} & \qw & \ctrl{1} & \ctrl{2} & \qw  & \targ & \targ & \targ & \ctrl{1} & \qw  & \ctrl{1} & \qw  & \ctrl{1} & \qw  & \ctrl{1} & \qw  & \qw  & \targ & \targ & \targ & \ctrl{1} & \qw  & \ctrl{1} & \qw  & \ctrl{1} & \qw  & \ctrl{1} & \ctrl{1} & \qw  \\
 & \ctrl{-1} & \ctrl{-1} & \ctrl{2} & \qw  & \ctrl{2} & \qw  & \qw  & \qw  & \targ & \targ & \targ & \targ & \targ & \targ & \targ & \qw  & \qw  & \qw  & \qw & \qw  & \targ & \targ & \targ & \targ & \targ & \targ & \targ & \ctrl{1} & \qw  & \qw  & \qw  & \qw  & \targ & \targ & \targ & \targ & \targ & \targ & \targ & \targ & \qw  \\
\lstick{|0\>}&  \qw  & \qw  & \qw  & \qw  & \qw  & \qw  & \ctrl{-2} & \qw  & \qw  & \qw  & \qw  & \ctrl{-1} & \qw  & \qw
 & \qw  & \targ & \ctrl{1} & \qw  & \ctrl{-2} & \qw  & \qw  & \qw  & \qw  & \ctrl{-1} & \qw  & \qw  & \qw  & \targ & \ctrl{1} & \qw  & \ctrl{-2} & \qw  & \qw  & \qw  & \qw  & \ctrl{-1} & \qw & \qw
 & \qw & \qw & \qw  \\
\lstick{|0\>}  & \qw  & \qw  & \targ & \targ & \targ & \ctrl{-3} & \qw  & \ctrl{-3} & \qw  & \ctrl{-2} & \qw  & \qw  &
\qw  & \ctrl{-2} & \qw  & \qw  & \targ & \ctrl{-3} & \qw  & \ctrl{-3}
& \qw  & \ctrl{-2} & \qw  & \qw  & \qw  & \ctrl{-2} & \qw  & \qw  & \targ & \ctrl{-3} & \qw  & \ctrl{-3} & \qw  & \ctrl{-2} & \qw  & \qw  &
\qw  & \ctrl{-2} & \qw & \qw & \qw 
}
}\label{subfig:seq V2 decomp}}
\caption{\label{fig:decomp seq}Decomposition of isometries corresponding to an optimal sequential strategy for $N=2$. We apply $V^{(2)\prime}$ instead of $V^{(2)}$ to achieve the maximal QFI with fewer CNOT gates.}
\end{figure}

Now we present numerical results of the circuit implementation of an optimal sequential strategy. The decomposition of ismometries is implemented using the Mathematica package UniversalQCompiler \cite{Iten19Universal} based on the method described above. We consider the phase estimation with amplitude damping noise described by Eq.~(\ref{eq:AD noise}) and take $N=2$, $\phi=\pi/2$, and $p=0.5$. The circuits implementing $V^{(1)}$ and $V^{(2)\prime}$ are illustrated in Figure~\ref{fig:decomp seq}. The state preparation $V^{(1)}$ requires $1$ CNOT gate and the intermediate control operation $V^{(2)\prime}$ requires $40$ CNOT gates.

\threesubsection{Optimal causal superposition strategy} \label{app:implementation of sup}
A causal superposition strategy for estimating $N$ channels can be implemented by an $N!$-dim quantum control system entangled with $N!$ sequential strategies of applying the channels:
\begin{equation} \label{app:eq:causal superposition strategy}
    P = \dyad{P}\ \mathrm{for}\ |P\>=\sum_{\pi\in S_N} |P^{\pi}\>|\pi\>_C,
\end{equation}
where $\{|\pi\>_C\}$ forms an orthonormal basis of the Hilbert space $\map H_C$ of the control system, and each $P^{\pi}=\dyad{P^\pi}$ is a sequential strategy. Once we obtain an optimal causal superposition strategy by applying Algorithm \ref{alg_optimal_strategy}, we can apply the circuit decomposition for each sequential strategy in the superposition. 

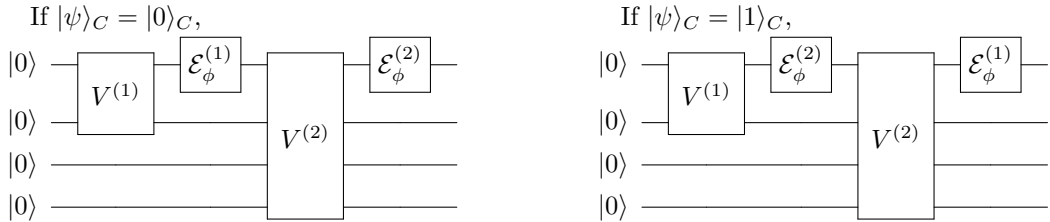
\begin{figure} [!htbp]
    \centering
\mbox{
\Qcircuit @C=1em @R=.7em {
& \mbox{If $|\psi\>_C = |0\>_C$,}  & & & & \\
\lstick{|0\>} & \multigate{1}{V^{(1)}} & \gate{\map E_\phi^{(1)}} & \multigate{3}{V^{(2)}} & \gate{\map E_\phi^{(2)}} & \qw \\
\lstick{|0\>} & \ghost{V^{(1)}}& \qw & \ghost{V^{(2)}} & \qw & \qw \\
\lstick{|0\>} & \qw & \qw & \ghost{V^{(2)}} & \qw & \qw \\
\lstick{|0\>} & \qw & \qw & \ghost{V^{(2)}} & \qw & \qw 
}
} \hspace{6em}
\mbox{
\Qcircuit @C=1em @R=.7em {
& \mbox{If $|\psi\>_C = |1\>_C$,} & & & & \\
\lstick{|0\>} & \multigate{1}{V^{(1)}} & \gate{\map E_\phi^{(2)}} & \multigate{3}{V^{(2)}} & \gate{\map E_\phi^{(1)}} & \qw \\
\lstick{|0\>} & \ghost{V^{(1)}}& \qw & \ghost{V^{(2)}} & \qw & \qw \\
\lstick{|0\>} & \qw & \qw & \ghost{V^{(2)}} & \qw & \qw \\
\lstick{|0\>} & \qw & \qw & \ghost{V^{(2)}} & \qw & \qw 
}
}
    \caption{Sequences of isometries corresponding to each sequential order in the causal superposition for $N=2$. The first qubit of the circuit is the system qubit, and the query order of two identical channels $\map E_\phi^{(1)}$ and $\map E_\phi^{(2)}$ is entangled with the state of the control qubit $|\psi\>_C$. When $|\psi\>_C$ is a superposition of the two states shown in the figure, the causal order is also in a superposition given by Eq.~(\ref{app:eq:causal superposition strategy}).}
    \label{fig:V1 and V2 sup}
\end{figure}

As a concrete example, we again take $N=2$, $\phi=\pi/2$, and $p=0.5$ for the amplitude damping noise and present numerical results of the circuit implementation of an optimal causal superposition strategy. As illustrated in Figure~\ref{fig:V1 and V2 sup}, we use the qubit $|\psi\>_C$ to coherently control which sequential order is executed. Due to the permutation invariance of the optimal strategy, we can simply control the query order of the identical channels while fixing $V^{(1)}$ and $V^{(2)}$ for all sequential orders. In view of this, generally we can use a $(2N-1)$-quantum SWITCH to control the order of $N$ channels $\map E_\phi$ and $N-1$ intermediate control operations $V^{(i)}$.

\begin{figure} [!htbp]
\centering
\subfigure[Decomposition of $V^{(1)}$. For simplicity the angles of single-qubit rotation gates are not depicted.]{\mbox{
    \Qcircuit @C=1em @R=.7em {
\push{\rule{20em}{0em}} & \lstick{|0\>} & \gate{R_y} & \gate{R_z} & \ctrl{1} & \qw & \push{\rule{20em}{0em}}  \\
\push{\rule{20em}{0em}} & \lstick{|0\>} & \qw & \qw & \targ & \qw & \push{\rule{20em}{0em}}
}
}\label{subfig:sup V1 decomp}}\\
\subfigure[Decomposition of $V^{(2)}$. For simplicity single-qubit gates, which might be required in addition to CNOT gates, are not depicted.]{\mbox{
    \Qcircuit @C=.4em @R=.7em {
& \qw  & \targ & \targ & \qw  & \ctrl{3} & \qw  & \targ & \targ & \targ & \ctrl{1} & \qw  & \ctrl{1} & \qw  & \ctrl{1} & \qw & \ctrl{1} & \ctrl{2} & \qw  & \ctrl{3} & \targ & \targ & \targ & \qw  & \qw  & \qw  & \targ & \ctrl{1} & \qw  & \ctrl{1} & \targ & \qw  & \targ & \targ & \qw  \\
& \qw  & \ctrl{-1} & \ctrl{-1} & \ctrl{2} & \qw  & \ctrl{2} & \qw & \qw  & \qw  & \targ & \targ & \targ & \targ & \targ & \targ & \targ & \qw  & \qw  & \qw  & \qw  & \qw  & \qw  & \ctrl{1} & \qw  & \ctrl{2} & \ctrl{-1} & \targ & \targ & \targ & \qw  & \targ & \ctrl{-1} & \ctrl{-1} & \qw  \\
\lstick{|0\>}& \qw  & \qw  & \qw  & \qw  & \qw  & \qw  & \qw  & \ctrl{-2} & \qw  & \qw & \qw  & \qw  & \ctrl{-1} & \qw  & \qw & \qw  & \targ & \ctrl{1} & \qw  & \qw  & \ctrl{-2} & \qw  & \targ & \ctrl{1} & \qw & \qw  & \qw  & \ctrl{-1} & \qw  & \ctrl{-2} & \qw  & \qw & \qw  & \qw  \\
\lstick{|0\>}& \qw  & \qw  & \qw  & \targ & \targ & \targ & \ctrl{-3} & \qw & \ctrl{-3} & \qw  & \ctrl{-2} & \qw  & \qw  & \qw  & \ctrl{-2} & \qw  & \qw  & \targ & \targ & \ctrl{-3} & \qw  & \ctrl{-3} & \qw  & \targ & \targ & \qw  & \qw  & \qw  & \qw  & \qw &
\ctrl{-2} & \qw & \qw  & \qw    
}
}\label{subfig:sup V2 decomp}}
\caption{\label{fig:decomp sup}Decomposition of isometries corresponding to one sequential order of an optimal causal superposition strategy for $N=2$. We have already taken advantage of the freedom to choose a $V^{(2)}$ implemented by fewer CNOT gates.}
\end{figure}
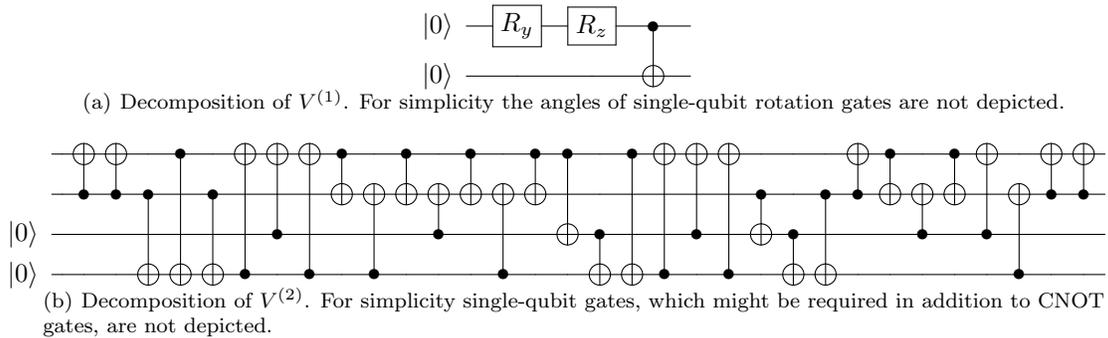

By further decomposition it turns out that each sequential branch requires one CNOT gate for state preparation $V^{(1)}$ and 32 CNOT gates for the intermediate control $V^{(2)}$, as illustrated in Figure \ref{fig:decomp sup}.

\subsubsection{Optimal protocols by variational circuits} \label{sec:optimal protocols vc}

We have shown how to design quantum circuits for strictly optimal quantum metrology. However, the exponential circuit complexity imposes strong restrictions on the experimental implementation. To circumvent the exponential overhead, we discuss two alternative approaches to the design of (nearly) optimal protocols based on variational circuits, which can often achieve a better gate count, possibly at the cost of a lower estimation accuracy. Here we only focus on optimizing sequential strategies, while these methods can also be similarly applied to identify optimal causal superposition strategies. 

\threesubsection{Variational optimization of the Fisher information}
The first approach is independent of Algorithm \ref{alg_optimal_strategy}, and thus does not require finding an optimal strategy by SDP. Instead, we employ a variational quantum circuit to maximize the classical Fisher information of the measurement statistics of the output state, by optimizing probe states, intermediate control operations and the measurement basis. Variational algorithms have been used to search for optimal probe states/measurements in both single-parameter \cite{Koczor_2020,Yang2020,Kaubruegger19PRL,Kaubruegger21PRX,Ma21IEEE,Beckey22PRRvariational} and multiparameter quantum metrology \cite{Meyer2021,Kaubruegger23PRXQuantum,Le2023variational,Cimini2024}. The variational quantum circuit used here is similar to the one in Ref.~\cite{Altherr2021PRL}, where not only the probe state and measurement but also the control operation is optimized.

\begin{figure} [!htbp]
\centering
\subfigure[Variational circuit ansatz.]{\includegraphics[height=0.18\textwidth]{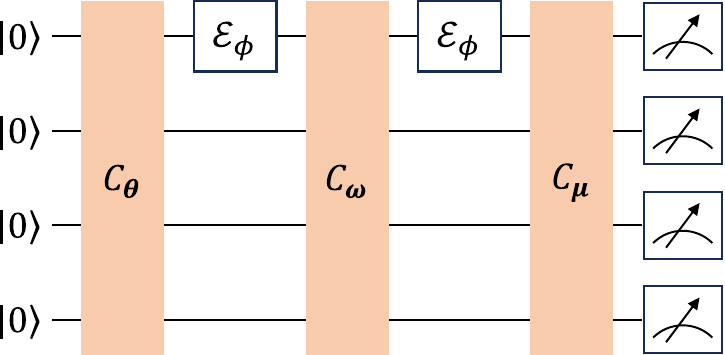}\label{subfig:variational_circuit_CFI}} \qquad
\subfigure[Variational block.]{\includegraphics[height=0.18\textwidth]{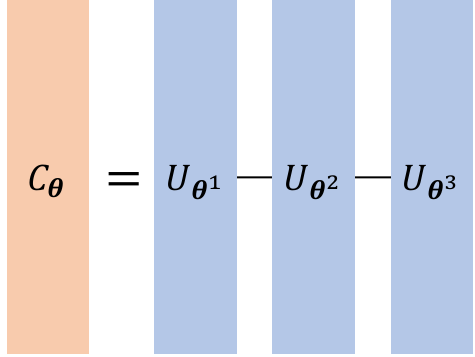}\label{subfig:variational_block}} \qquad
\subfigure[Variational layer.]{\includegraphics[height=0.18\textwidth]{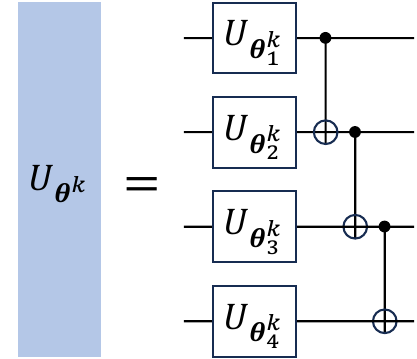}\label{subfig:variational_layer}}
\caption{\label{fig:variational_ansatz_CFI}Variation circuit ansatz for maximizing the classical Fisher information for the estimation of $N=2$ channels $\mathcal E_\phi$. In the example illustrated here, the circuit for $n_q=4$ qubits comprises three variational blocks $C_{\boldsymbol \theta},C_{\boldsymbol \omega},C_{\boldsymbol \mu}$, corresponding to the probe state, control operation and measurement basis. Each block consists of $n_l=3$ variational layers, and each layer contains arbitrary single-qubit unitary gates (determined by three angle parameters in ZYZ decomposition \cite{Nielsen_Chuang_2010}) and a sequence of CNOT gates connecting neighboring qubits.}
\end{figure}

A main advantage of this method is the flexibility of the variational ansatz, as we have the freedom to choose the number of qubits and the arrangement of variational gates. It is thus possible to start with a relatively small number of qubits and gates, and gradually increase the number until the metrological performance meets the requirement. As a proof-of-principle application, we choose a simple variational ansatz illustrated in Figure~\ref{fig:variational_ansatz_CFI}. For the estimation of $N$ channels $\mathcal E_\phi$, we introduce a variational circuit with $n_q$ qubits and $N+1$ blocks, and each block comprises $n_l$ layers. The whole circuit contains $(N+1)n_l(n_q-1)$ CNOT gates
and $3(N+1)n_ln_q$ variables. Note that we only allow for single-qubit gates and CNOT gates, which is thus comparable to the optimal protocol by the comb decomposition. The cost function is the classical Fisher information defined by Eq.~(\ref{eq:CFI}), and the measurement is performed in the computational basis. Note that computing the Fisher information requires the calculation of derivatives with respect to $\phi$, which can be either classically performed or evaluated on a quantum computer using the parameter-shift rule \cite{Li17PRLhybrid,Mitarai18PRAquantum,Schuld19PRAevaluating}.

\threesubsection{Variational circuit compilation of quantum combs}
In the second approach, we integrate Algorithm \ref{alg_optimal_strategy} with variational algorithms, to leverage both the guaranteed optimality of Algorithm \ref{alg_optimal_strategy} and the practical performance of variational circuits. Based on the output of Algorithm \ref{alg_optimal_strategy}, we also propose a variational approach for the circuit implementation of the optimal strategy. To this end, we first decompose the Choi operator output by Algorithm \ref{alg_optimal_strategy} into a sequence of isometries as described in Section \ref{sec:circuit decomposition theory}. For each isometry from $m$ to $n$ qubits, we apply the variational circuit compiling approach proposed by Ref.~\cite{Khatri2019quantumassisted} to optimize a variational circuit approximating the isometry, by introducing $n-m$ ancilla qubits. Denoting the target isometry by $V_{(\mathrm{target})}$ and the variational circuit by $U(\boldsymbol \theta)$, the cost function is taken as
\begin{equation}
    \mathrm{Cost}(\boldsymbol\theta)=1-\frac{1}{4^{m}}\left\lvert\Tr\left[V_{(\mathrm{target})}^\dagger U(\boldsymbol \theta)\right]\right\rvert^2,
\end{equation}
which is the infidelity between Choi states of $V_{(\mathrm{target})}$ and $U(\boldsymbol \theta)$, and also has a close relation to the average gate fidelity over Haar random input states \cite{Horodecki99PRA,Nielsen02PLA}. Note that $\mathrm{Cost}(\boldsymbol\theta)=0$ if and only if $V_{(\mathrm{target})}$ and $U(\boldsymbol \theta)$ are equivalent up to an irrelevant global phase factor. The evaluation of the cost function can be either performed by classical simulation, or implemented on a quantum computer using the Hilbert-Schmidt test \cite{Khatri2019quantumassisted}.

\begin{figure} [!htbp]
    \centering
    \includegraphics[width=0.3\textwidth]{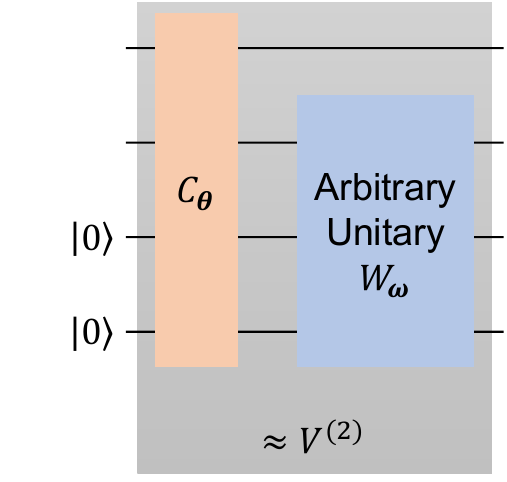}
    \caption{Variational ansatz for compiling an isometry $V^{(2)}$. Two blocks $C_{\boldsymbol \theta}$ and $W_{\boldsymbol \omega}$ are optimized in an alternating way.}
    \label{fig:variational_ansatz_compiling}
\end{figure}

As a concrete example, assume that by Algorithm \ref{alg_optimal_strategy} we have already obtained an optimal sequential strategy for $N=2$ channels $\mathcal E_\phi$, and then we decompose it into isometries $V^{(1)}$ and $V^{(2)}$ as in Figure~\ref{fig:V1 and V2 seq}. Implementing $V^{(1)}$ only requires one CNOT gate, so we will only focus on the variational compilation of the two to four qubit isometry $V^{(2)}$. The variational circuit ansatz $U(\boldsymbol \theta, \boldsymbol \omega)=(I\otimes W_{\boldsymbol \omega})C_{\boldsymbol \theta}$ for approximating $V^{(2)}$ is illustrated by Figure~\ref{fig:variational_ansatz_compiling}. $C_{\boldsymbol \theta}$ is taken to have the same structure as the variational block in Figure~\ref{fig:variational_ansatz_CFI}\protect\subref{subfig:variational_block} with $n_l$ layers, therfore comprising $12n_l$ variables and $3n_l$ CNOT gates. $W_{\boldsymbol \omega}$, which is an arbitrary three-qubit unitary parameterized by $4^3-1=63$ variables (angles in Pauli string rotations), acts on the ancilla qubits without affecting the QFI of the output state. Hence, $\boldsymbol \omega$ represents the free parameters for better optimization, and there is no need to implement $W_{\boldsymbol \omega}$ in the optimized protocol at last. We can thus perform alternating minimization of $\mathrm{Cost}(\boldsymbol\theta, \boldsymbol\omega)$ over $C_{\boldsymbol \theta}$ and $W_{\boldsymbol \omega}$, as described in Algorithm \ref{alg_alternating_optimization}. For each iteration $k$ of the outer loop, we first tune $\boldsymbol \theta$ for fixed $\boldsymbol \omega$ until convergence, and then tune $\boldsymbol \omega$ for fixed $\boldsymbol \theta$ until convergence. Finally, the optimized $C_{\boldsymbol \theta}$ itself can replace $V^{(2)}$ in the optimal protocol.
\SetKwComment{Comment}{/* }{ */}
\begin{algorithm}[!htbp]
\caption{Alternating optimization of $\mathrm{Cost}(\boldsymbol\theta, \boldsymbol\omega)$.}\label{alg_alternating_optimization}
\DontPrintSemicolon
Initialize $\boldsymbol \theta=\boldsymbol \theta^{(0)}$ and $\boldsymbol \omega=\boldsymbol \omega^{(0)}$ \\
\KwIn{maximum number of iterations $K$} 
\For{$k=1,\dots,K$}
{
    $\boldsymbol \theta^{(k)} \gets \argmin_{\boldsymbol \theta}\mathrm{Cost}(\boldsymbol\theta, \boldsymbol\omega^{(k-1)})$ \Comment*[r]{By gradient descent until convergence}
    $\boldsymbol \omega^{(k)} \gets \argmin_{\boldsymbol \omega}\mathrm{Cost}(\boldsymbol\theta^{(k)}, \boldsymbol\omega)$ \Comment*[r]{By gradient descent until convergence}
}
\end{algorithm}    

This approach also applies to the case where we have the Choi operator description (quantum comb) of the noisy physical process encoding the parameter of interest, but the necessary control is not achievable on the sensing platform directly and the noise cannot be readily simulated by quantum circuits. For example, when a finite-dimensional system goes though an $N$-step process with non-Markovian noise resulting from the interaction of the system and an infinite-dimensional environment, the quantum comb of the $N$-step process is an operator only on the system spaces at different time steps and there is no need to simulate the infinite-dimensional environment directly.

\subsubsection{Comparison of different approaches}
\begin{figure} [!htbp]
    \centering
    \includegraphics[width=0.7\textwidth]{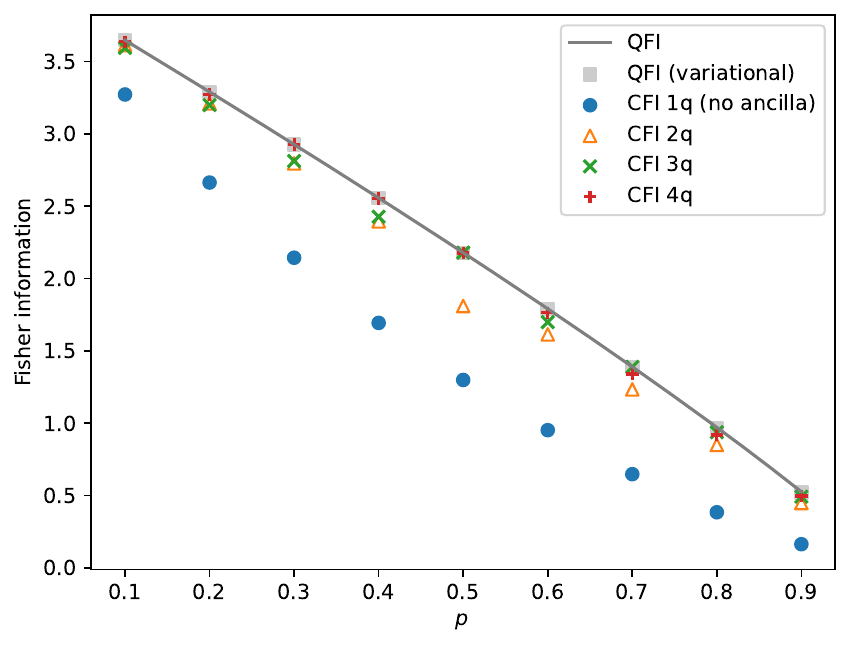}
    \caption{Comparison between the Fisher information obtained by different methods of optimizing strategies. The strictly optimal QFI (attained by the strategy yielded by Algorithm \ref{alg_optimal_strategy}) is plotted as a gray line, which serves as an upper bound on the Fisher information that any protocol can achieve. The light gray squares represent the QFI achieved by the variational compilation of the optimal strategy. The other markers represent the optimized CFI by variational circuits with $n_q=1,2,3,4$ qubits respectively.}
    \label{fig:comparison qfi cfi}
\end{figure}

We test the performances of these three different approaches introduced above for a specific example. We take $N=2$, $\phi=\pi/2$ for the phase estimation with the amplitude damping noise characterized by Eq.~(\ref{eq:AD noise}). In Figure~\ref{fig:comparison qfi cfi}, for different noise strength $p$, we compare the optimal QFI attained with the strategy output by Algorithm $\ref{alg_optimal_strategy}$, the QFI achieved by the variational compiling of the optimal strategy, and the classical Fisher information (CFI) optimized by variational circuits. In this example, the comb decomposition of the optimal strategy yields a zero to two qubit isometry $V^{(1)}$ and a two to four qubit isometry $V^{(2)}$. Therefore, the variational compiling of $V^{(2)}$ also requires four qubits, and we take $n_l=4$ layers for $C_{\boldsymbol \theta}$ in the numerical experiment. For the variational optimization of CFI, we take $n_l=3$ layers for each block (nine layers in total) and test the performance for the number of qubits $n_q=1,2,3,4$. We do not observe significant improvement by further increasing the number of layers.

While an exact decomposition of $V^{(2)}$ typically requires dozens of CNOT gates, remarkably, the variational compiling of $V^{(2)}$ has almost attained the optimality with only $12$ CNOT gates, which underscores the practical potential of combining Algorithm \ref{alg_optimal_strategy} and the variational approach. In terms of the variational optimization of the CFI by variational circuits, increasing the number of qubits mostly tends to improve the Fisher information, albeit not necessarily. When $n_q=4$, the CFI is also very close to optimal (slightly worse than the QFI obtained by variational compiling). In some cases, e.g., when $p=0.1$, $n_q=2$ qubits (i.e., one ancilla qubit) already result in a nearly optimal CFI, which could save the qubit cost and facilitate the experimental implementation.

\subsection{Experimental implementation of optimal protocols}

In this subsection we briefly discuss how our theoretical framework relates to the optimal sequential and quantum SWITCH protocols that have been implemented in experiments. The optimal protocols for other specific tasks and experimental setups can be implemented similarly.

\subsubsection{Optimal sequential protocol} 

The optimal sequential protocol via adaptive controls \cite{Yuan2015PRL} has been applied in phase estimation in an optical platform to recover the Heisenberg limit for non-commuting dynamics \cite{Hou2019PRL}. The initial state representing the polarization of the heralded photon can be prepared using a half-wave plate (HWP) and a
quarter-wave plate (QWP). The evolution operator $U_\phi = e^{-iH(\phi)t}$ with $H(\phi) = \sin(\phi) X + \cos(\phi) Z$ is realized by an adjustable phase
plate realized with a Soleil-Babinet Compensator and the controls can also be realized as arbitrary unitary evolution using QWPs and HWPs. Using HWP, QWP, polarizing beam splitter (PBS) and single-photon detectors, projective measurement can be performed along an arbitrary direction. 

Since the system evolution for different $\phi$ do not commute, a sequential protocol without control can only achieve the QFI $4 \,\mathrm{sin}^2 T$ for total evolution time $T$. Using the optimal probe state $\mathrm{cos}\, \hat{\phi} |0\rangle + \mathrm{sin} \,\hat{\phi} |1\rangle $ and adaptive controls $U_c = U^{\dagger}_\phi$, the Heisenberg limit can be recovered. Note that the above discussion is for the case of unitary dynamics. For generic noisy quantum evolution, the optimal QFI can be obtained via the SDP in Theorem \ref{thm_qfi_sdp}. The optimal initial state and controls can then be obtained numerically by decomposing the the output of Algorithm \ref{alg_optimal_strategy} into quantum circuits using the techniques introduced in Section \ref{sec:circuit decomposition theory}. 

\subsubsection{Optimal quantum SWITCH protocol}

The quantum SWITCH strategy achieving the super-Heisenberg scaling \cite{Zhao2020PRL} has been realized experimentally in Ref.~\cite{Yin2023NP}, where the system is chosen as a single photon whose polarization and transverse spatial modes are treated as the control and target systems respectively. The control system is initialized as an equal superposition of horizontal and vertical polarization $(|H\> + |V\>)/\sqrt{2}$ by a PBS and a HWP, which is then used to generate superposition of paths through a Mach-Zehnder interferometer. The quantum SWITCH has also been experimentally realized in Refs.~\cite{cao2022PRL,Goswami2018PRL} in optical platforms. For generic quantum SWITCH strategies, our framework can be used to calculate the optimal QFI via SDP and the optimal initial state using Algorithm \ref{alg_optimal_strategy}.

\section{Conclusion and Future perspectives} \label{sec:conclusions}
In this tutorial, a systematic approach for identifying and designing the optimal strategy with maximal quantum Fisher information has been presented for the estimation of a single parameter in the local regime. Possible extensions include quantum metrology under limited resources, Bayesian quantum parameter estimation, quantum parameter estimation with finite samples, and multiparameter quantum estimation. Some of these have been considered in recent works \cite{meyer2023quantum,zhou2024strict,kurdzialek2024quantummetrologyusingquantum,liu2024efficienttensornetworkscontrolenhanced}.

The optimal strategy may require significant resources in terms of the size of ancillary systems, depth of circuit required for the preparation of optimal probe states, implementation of optimal control and measurement. Under limited resources, it is crucial to develop strategies that optimize the use of limited resources while still achieving high precision. This involves techniques such as resource-efficient state preparation, adaptive measurements, and optimization algorithms tailored to resource limitations. Recent works have investigated the limits of noisy quantum metrology with restricted ancilla and control \cite{zhou2024limitsnoisyquantummetrology,liu2024heisenberglimitedquantummetrologyancilla}, and optimization algorithms under certain resource constraints have also been developed \cite{kurdzialek2024quantummetrologyusingquantum,liu2024efficienttensornetworkscontrolenhanced}. Nevertheless, general approaches to resource-constrained metrology remain lacking.

Systematic approaches that can provide optimal strategies with minimal cost in the Bayesian regime are also desired. Bayesian estimation provides a framework for incorporating prior knowledge and updating estimates based on observed data, which is closely connected to the quantum parameter estimation with finite samples \cite{meyer2023quantum,Chiribella_2012,Rubio_2018,Rubio_2019,Bavaresco24PRRdesigning}. Identification of optimal strategies in the Bayesian regime requires the development of statistical methods that can handle the inherent quantum fluctuations and sampling uncertainties. Additionally, the trade-off between the number of samples and the achievable precision needs to be investigated to optimize the allocation of resources in practical applications.

Another significant extension is the consideration of scenarios where multiple parameters need to be estimated simultaneously. Multiparameter quantum estimation arises in various applications, such as quantum imaging, quantum gyroscopes and quantum tomography, where information about multiple parameters is desired. The challenge in multiparameter quantum estimation lies in the trade-off between the precision of individual parameter estimates and the correlations between them. The study of multiparameter quantum estimation is an active area of research \cite{HongzhenPra,HongzhenPRL,GillM00,Lu2021,Suzuki2016,Sidhu2021,Nagaoka1,ALBARELLI2020126311,Albarelli20PRL,Federico2021,Albarelli22PRX,Hou20minimal,HouSuper2021,Houeabd2986}, with many problems remaining open. 
Due to the incompatibility between multiple parameters, the QFI matrix may no longer provide the most insight about the best achievable precision, and there has been increasing number of works utilizing other metrological bounds tighter than the multiparameter SLD quantum Cram\'{e}r-Rao bound \cite{Rafal2020,Conlon2021,Hayashi2023tightcramerraotype,hayashi2024finding}.

\medskip
\textbf{Acknowledgements} \par 
The authors thank Stanisław Kurdziałek for providing the code for computing QFI bounds in Ref.~\cite{kurdzialek2023using}. This work was supported by the National Natural Science Foundation of China via Excellent Young Scientists Fund (Hong Kong and Macau) Project 12322516, Guangdong Basic and Applied Basic Research Foundation (Project No.~2022A1515010340), Ministry of Science and Technology, China (MOST2030 with Grant No 2023200300600 and 2023200300603), the Guangdong Provincial Quantum Science Strategic Initiative (Grant No.GDZX2303007), and the Hong Kong Research Grant Council (RGC) through the Early Career Scheme (ECS) grant 27310822 and the General Research Fund (GRF) grant (17303923, 14307420, 14308019, 14309022, and 14309223). 
The numerical results are obtained via the Python package CVXPY \cite{diamond2016cvxpy,agrawal2018rewriting} with the solver MOSEK \cite{mosek}, the Python package PennyLane \cite{bergholm2022pennylane}, and the Mathematica package UniversalQCompiler \cite{Iten19Universal}.

\medskip
\textbf{Conflict of Interest} \par 
The authors declare no conflict of interest.
%
\medskip
\bibliographystyle{MSP}
\bibliography{reference}



\end{document}